%% file: ms.tex
\documentclass[10pt, letterpaper]{article}
\usepackage[utf8]{inputenc}
\usepackage[T1]{fontenc}
\usepackage[margin=1in]{geometry}
\usepackage[dvipsnames]{xcolor}
\usepackage{amsmath,amssymb,amsfonts}
\usepackage{graphicx}
\usepackage{booktabs}
\usepackage{natbib}
\usepackage{hyperref}
\hypersetup{
  colorlinks=true,
  linkcolor=BrickRed,
  citecolor=ForestGreen,
  urlcolor=NavyBlue,
}
\usepackage{showyourwork}

\title{How long can you trust a Starlink TLE? \\
       An empirical comparison of SGP4 and high-fidelity propagation
       against operator-updated truth across a megaconstellation}

\author{Dimitrije Jankovic \\
        \small Independent researcher --- astro-tools \\
        \small \href{https://github.com/astro-tools/paper-tle-divergence-atlas}{github.com/astro-tools/paper-tle-divergence-atlas}}

\date{2026-05-18}

\begin{document}

\maketitle

\begin{abstract}
\noindent
We characterise the position-error behaviour of public Two-Line
Element (TLE) propagation against operator-updated truth on the
Starlink megaconstellation, sweeping $24{,}641$ next-TLE-truth
pairs across $501$ satellites stratified by altitude shell
($540$, $550$, $560$~km) and platform generation (v1.0, v1.5,
v2-mini) over April $2026$. Each pair is propagated with both SGP4
and NASA's General Mission Analysis Tool (GMAT) at high fidelity
(EGM2008 $70\!\times\!70$, NRLMSISE-00 drag, Sun and Moon
third-body gravity, conical-shadow SRP), and the two predictions
are compared against the operator's subsequently released ephemeris
evaluated as proxy truth. Three findings emerge. First, position
error follows a per-cell power law
$\lVert\Delta\mathbf{r}(\Delta t)\rVert \approx A\,\Delta t^{k}$
whose fitted exponents lie in $(1,\,2)$ on every populated v2-mini
cell and on the high-fidelity v1.x propagator at $540$ and $560$~km,
while the SGP4 v1.x cells and the high-fidelity v1.x cell at
$550$~km are sub-linear ($k \lesssim 1$); the corpus is therefore
consistent with a mixture of mean-motion bias and constant unmodelled
in-track acceleration whose relative weight varies by cohort; pooled L$_{2}$ medians grow from
$\sim 1$~km at the $6$~h horizon (just beyond the operator's
$\sim 4$-hour update interval) to $\sim 38$~km (SGP4) /
$\sim 76$~km (high-fid) at $7$~d. Second, the high-fidelity
propagator initialised from a public TLE does not improve over
SGP4 at any of the four staleness horizons sampled; SGP4 wins on
$\sim 65$--$75\%$ of pairs depending on horizon --- with the
v2-mini cohort at long $\Delta t$ the one (cohort $\times$ horizon)
regime where high-fidelity overtakes SGP4 on a majority of pairs,
visible at both populated shells. The bulk negative result
decomposes into three mechanisms --- at-epoch operator-OD
residual dominance, SGP4-vs-SGP4 truth-construction kernel
alignment, and preferential spacecraft-property bias amplification
on the high-fidelity arm. Third, an exploratory regression of the
per-satellite SGP4 staleness coefficient against F10.7 returns a
positive slope at one of three altitude shells ($560$~km) that
clears conventional significance, with the other two shells slope-
ambiguous on the present $30$-day, moderate-activity, $\sim 17$~sfu
window; we read the H3 evidence as direction-consistent with the
LEO density-gradient expectation but \emph{not} a calibrated
F10.7-modulation measurement on this corpus. Practitioners
should regard public Starlink TLEs as adequate inside the operator's
update cadence, treat the per-cell power-law fits as cohort-
resolved staleness budgets, and avoid investing in
high-fidelity propagation from public-TLE inputs unless improved
initial states are also available. All code, the locked corpus,
the resumable sweep manifest, and the rendered figures are
released together under the MIT license.
\end{abstract}

\section{Introduction}
\label{sec:introduction}

The orbital population in low Earth orbit (LEO) has shifted from a sparse
collection of single-platform missions to large operator-fielded
megaconstellations within a single decade, with SpaceX's Starlink alone
accounting for the majority of active payloads below 600~km
\citep{mcdowell2020}. The public Two-Line Element (TLE) data product,
derived by the United States Space Force from radar and optical
observations and propagated with the analytic SGP4 model, remains the
de facto state input for academic conjunction-analysis pipelines,
commercial space-situational-awareness toolchains, and independent
observers --- none of whom has access to operator-internal
orbit-determination state or onboard GNSS solutions. The operational
character of TLE delivery has, however, evolved with the constellation
era: Starlink TLEs publish at roughly six updates per satellite per day,
an order of magnitude faster than the once-per-day cadence assumed by
the prior TLE-accuracy literature \citep{vallado2006,vallado2012}. The
operationally relevant staleness question for a present-day TLE consumer
is therefore measured in hours, not weeks --- the boundary between fresh
and stale lies within a single operator-update interval, and an error
budget framed against propagation horizons of seven days
under-represents the regime in which most downstream tools actually
operate.

The next-TLE-as-truth methodology developed by \citet{vallado2006},
\citet{vallado2012}, and \citet{vallado_crawford2008} established the
analytic apparatus for measuring SGP4 accuracy from the operator's own
subsequently released ephemeris. Those studies were necessarily small
in scale: a handful of well-tracked individual platforms, observed at
the update cadences typical of their era. The constellation-scale,
multi-generation, public-data regime that emerged with Starlink opens
three questions that the original analysis was not positioned to
address. First, how the SGP4 propagation error scales with time since
epoch across orbital regime and platform generation in the
consumer-relevant horizon range. Second, whether the high-fidelity
force models routinely available in modern open-source mission-analysis
tools materially improve accuracy when initialised from a TLE whose own
orbit-determination residual is already non-zero at epoch. Third,
whether short-window solar-flux modulation of thermospheric density is
detectable in the empirical error-growth coefficients of a confined
operational fleet.

We address the three questions by sweeping 24{,}641
starting-TLE\,/\,next-TLE pairs across 501 Starlink satellites,
stratified by altitude shell (540, 550, 560~km) and platform generation
(v1.0, v1.5, v2-mini) over the one-month window of April 2026. Each
pair is propagated both with SGP4 from the starting TLE and with a
high-fidelity propagation in NASA's General Mission Analysis Tool
(GMAT), and the two predictions are compared against the operator's
next-TLE position evaluated as proxy truth. We fit a per-cell power-law
staleness curve $\lVert\Delta\mathbf{r}(\Delta t)\rVert \approx
A\,\Delta t^{k}$ with satellite-level bootstrap uncertainties
(Hypothesis~H1, Figures~\ref{fig:sgp4-growth}
and~\ref{fig:hifi-growth}; parametrised in
Figure~\ref{fig:powerlaw-fits} and Table~\ref{tab:powerlaw}), test
pair-by-pair whether the high-fidelity propagator beats SGP4
(Hypothesis~H2, Figure~\ref{fig:propagator-scatter}), and regress the
per-satellite coefficient $A$ against the daily observed F10.7 flux in
the most-drag-sensitive shell (Hypothesis~H3,
Figure~\ref{fig:solar-modulation}). The data-preparation pipeline,
the locked corpus, the resumable manifest of every sweep run, and the
manuscript \LaTeX{} source are released together as an open-source
artefact under the MIT licence; the sweep output bundle is deposited
on Zenodo and fetched automatically by the manuscript build, so a
clean re-render does not require local installation of GMAT.
Section~\ref{sec:background} reviews the methodological lineage;
Section~\ref{sec:methods} specifies the corpus, force model, and
error metrics; the results, their interpretation, and the practical
consequences for downstream TLE consumers follow in
Sections~\ref{sec:results}--\ref{sec:conclusion}.

\section{Background and related work}
\label{sec:background}

SGP4 is the analytic averaged-element propagator paired with the TLE
format and standardised through the open-source reference
implementation distributed with \citet{vallado2006}. It expresses the
mean motion and Brouwer mean elements at epoch, applies analytic
secular corrections for $J_2$, $J_3$, and $J_4$ together with
long- and short-period perturbation series, and absorbs atmospheric
drag into a single fitted scalar $B^{\!\star}$ representing the mean
ballistic coefficient encountered during the orbit-determination
window. \citet{hoots2004} trace the development of the model family
inside the U.\,S.\ space-surveillance system, and
\citet[\S\,11.6]{vallado2013} gives the textbook formulation. Two
operationally relevant properties recur throughout the present
analysis: the state delivered by SGP4 lives in the True Equator,
Mean Equinox (TEME) frame --- not in a standard inertial frame ---
and the fitted $B^{\!\star}$ encodes the thermospheric density
encountered \emph{during} the orbit-determination window rather than
during the forward propagation that consumers subsequently perform.

The use of an operator's subsequently released TLE as a truth proxy
for prediction accuracy was introduced by \citet{vallado2006} alongside
the open-source SGP4 reference implementation, and was developed
methodologically by \citet{vallado2012} and applied to single-satellite
orbit determination by \citet{vallado_crawford2008}. In that approach
the position obtained by propagating $\mathrm{TLE}_{i}$ forward to a
later epoch $t_{j}$ is compared against
$\mathrm{SGP4}(\mathrm{TLE}_{j},\,\Delta t = 0)$, i.\,e.\ the operator's
subsequently released ephemeris evaluated at its own epoch. The proxy
is not a ground-truth measurement but a self-consistency check against
the operator's own orbit determination, and it inherits whatever
observation-fit residual is present in the next TLE. Its principal
merit for an open-data study is that it requires no privileged access
to tracking observations or operator-internal state vectors, which
permits constellation-scale application from a single archived TLE
history. The operator's orbit determination is itself smoothed over
a fit window of order $1$--$2$ days, so the resulting $B^{\!\star}$
and mean motion at epoch are averaged quantities rather than
instantaneous ones --- relevant later for the
F10.7-modulation interpretation in Section~\ref{sec:results}.

\subsection{Interpretational limits of next-TLE-as-truth}
\label{sec:background-truth-floor}

The next-TLE proxy is not a ground-truth measurement, and treating
it as one would overclaim the precision of the comparison. A
position error measured against $\mathrm{SGP4}(\mathrm{TLE}_{j},\,
\Delta t = 0)$ decomposes into three contributions:
\begin{equation}
  \label{eq:error-budget}
  \underbrace{\Delta\mathbf{r}_{\mathrm{obs}}(t_{j})}_{\text{what we measure}}
  \;=\;
  \underbrace{\boldsymbol{\Phi}(t_{j},\,t_{i})\,\Delta\mathbf{r}_{\mathrm{OD}}(t_{i})}_{\text{epoch-}t_{i}\text{ OD residual, propagated}}
  \;+\;
  \underbrace{\Delta\mathbf{r}_{\mathrm{prop}}(t_{i}\!\to\!t_{j})}_{\text{propagator error}}
  \;-\;
  \underbrace{\Delta\mathbf{r}_{\mathrm{OD}}(t_{j})}_{\text{epoch-}t_{j}\text{ OD residual}},
\end{equation}
where $\Delta\mathbf{r}_{\mathrm{OD}}$ denotes the operator's
observation-fit residual at the indicated epoch and $\boldsymbol{
\Phi}$ is the state-transition matrix of the chosen propagator over
$(t_{i},\,t_{j}]$. Throughout this paper, every reported error norm
$\lVert\Delta\mathbf{r}_{\mathrm{SGP4}}\rVert$ or
$\lVert\Delta\mathbf{r}_{\mathrm{hifi}}\rVert$ is the left-hand side
of Eq.~\eqref{eq:error-budget}; the propagator term in the middle
cannot be cleanly isolated from observation residuals at either
epoch using public TLEs alone.

The practical consequence is a regime split. At long staleness
($\Delta t = 7$~days, where Section~\ref{sec:results} reports
median observed errors of $40$--$80$~km) the OD-residual terms in
Eq.~\eqref{eq:error-budget} are at the percent level relative to
the observed total (we quantify the floor at order $1$~km
empirically below), and the propagator term dominates whether
evaluated with SGP4 or with a high-fidelity force model. At the
operationally relevant 6~h horizon, where
Section~\ref{sec:results} reports a pooled median observed error of
$\sim 1$~km, the OD-residual floor is the same order as the signal
itself.

No operator publishes an internal OD-residual figure for Starlink.
The three public quantifications closest to the question bracket
it from different sides, none of them an exact match.
\citet{lang2025} apply a third-party unscented batch filter to
maneuvering Starlinks and recover 24-hour position predictions
with RMS error below $3$~km. The commercial radar tracking service
described in the \citet{leolabs_eoportal} survey reports state
estimates better than $1$~km at time of estimation for
$\sim 95\%$ of LEO objects (and better than $200$~m for
$\sim 50\%$), without distinguishing Starlink from the broader
population. SpaceX's own narrative submission to the FCC on
positioning, navigation, and timing \citep{spacex_fcc_pnt}
discusses terminal-segment metric-precision positioning but does
not publish a satellite OD residual. None of these directly
answers Eq.~\eqref{eq:error-budget}: \citet{lang2025} and
\citet{leolabs_eoportal} characterise third-party OD pipelines
that observe the constellation \emph{less} precisely than the
operator does --- SpaceX has onboard GNSS on every satellite ---
so the operator-internal residual at $t_{j}$ is plausibly tighter
than the $\sim 1$~km outer bound those third-party pipelines
place on it; and the FCC submission targets user-equipment
positioning, not space-segment OD. We therefore measure the floor
directly with the diagnostic below.

We quantify the floor empirically by repeatedly propagating
$\mathrm{TLE}_{j-N}$ to $t_{j}$ with the same GMAT high-fidelity
configuration the main sweep uses (Section~\ref{sec:methods-forcemodel})
and comparing to $\mathrm{SGP4}(\mathrm{TLE}_{j},\,\Delta t = 0)$
for $N \in \{1,\,2,\,3,\,6\}$. To capture inter-cohort variation
rather than over-generalising from a single representative, the
diagnostic runs the densest-sampled corpus satellite in each of the
eight populated (altitude shell~$\times$~generation) cells of
Table~\ref{tab:corpus} --- 122--131 TLEs per sat, 3{,}121 arcs in
total after the 100~m SMA-jump maneuver filter of
Section~\ref{sec:methods-pairs}. Per-cohort medians at the shortest
arc ($N = 1$, $\Delta t \approx 4$--$5$~h) are listed in
Table~\ref{tab:truth-floor}. Two features of the table carry the
interpretation. First, every populated cohort sits in $0.81$--$1.37$~km
at the 5~h scale; the floor is, to within a factor of two,
\emph{cohort-independent} --- v2-mini is not measurably worse than
v1.5, and altitude separates the cohorts only weakly. Second, the
pooled $N = 1$ median of $1.11$~km (IQR $0.23$--$2.89$~km) already
exceeds the SGP4 pooled median at $\Delta t = 6$~h reported in
Section~\ref{sec:results}, which is the empirical statement of
floor limitation: even with a strictly better propagator
\emph{and} a strictly fresher starting TLE, the residual against
the proxy truth cannot drop below $\mathcal{O}(1)$~km regardless
of cohort, because that floor is built into the two TLE-derived
states it is constructed from. The 6~h headline reported below
should be read accordingly --- as a measurement constrained by the
precision of public TLE delivery, not by propagator skill, and
constrained \emph{uniformly} across the corpus.

\begin{table}[ht]
  \centering
  \caption{Empirical truth-floor diagnostic, per (altitude shell
  $\times$ generation) cohort. Each row is the densest-sampled
  corpus satellite in that cell, swept across
  $N \in \{1,\,2,\,3,\,6\}$ TLE offsets back from $\mathrm{TLE}_{j}$;
  the reported $|\Delta\mathbf{r}|$ is measured against
  $\mathrm{SGP4}(\mathrm{TLE}_{j},\,\Delta t = 0)$ at $t_{j}$. Only
  the shortest-arc median is tabulated for compactness; larger $N$
  trends are in the diagnostic JSON. All arcs survive the 100~m
  SMA-jump maneuver filter. The 540~km~$\times$~v2-mini cell is empty
  in the corpus (Table~\ref{tab:corpus}).}
  \label{tab:truth-floor}
  \begin{tabular}{lllrrrrr}
    \toprule
    Shell (km) & Gen & NORAD & TLEs & arcs ($N\!=\!1$) & $\Delta t$ (h) & median $|\Delta\mathbf{r}|$ (km) & IQR (km) \\
    \midrule
    540 & v1.0    & 48555 & 122 &  64 &  4.4 & 1.02 & 0.01--2.98 \\
    540 & v1.5    & 49766 & 125 & 123 &  4.8 & 1.17 & 0.23--4.10 \\
    540 & v2-mini & ---   & --- & --- &  --- & ---  & --- \\
    550 & v1.0    & 48306 & 106 &  60 &  4.0 & 0.94 & 0.12--4.01 \\
    550 & v1.5    & 53675 & 131 & 112 &  4.8 & 1.19 & 0.53--3.05 \\
    550 & v2-mini & 65384 & 127 & 123 &  4.8 & 1.11 & 0.01--2.75 \\
    560 & v1.0    & 48881 & 124 & 123 &  4.8 & 1.31 & 0.50--2.39 \\
    560 & v1.5    & 56495 & 129 & 128 &  4.8 & 0.81 & 0.23--1.90 \\
    560 & v2-mini & 63926 & 125 & 111 &  4.8 & 1.37 & 0.01--3.64 \\
    \midrule
    \multicolumn{4}{l}{\textit{Pooled across cohorts}} & 844 & --- & 1.11 & 0.23--2.89 \\
    \bottomrule
  \end{tabular}
\end{table}

The constellation-era context is set by two recent results.
\citet{mcdowell2020} documents the shift in the LEO satellite
population during the first phase of Starlink deployment and
quantifies the resulting change in object density at the operational
shells. \citet{baruah2024} models the loss of 38 Starlink v1.5
satellites during the February 2022 moderate geomagnetic storm using
a shark-fin ram area of 4.48\,m$^{2}$ at $C_{D} = 1$, and provides the
per-generation drag cross-section anchor reused in
Section~\ref{sec:methods-spacecraft-props}. Peer-reviewed quantitative
work on public-TLE accuracy at constellation scale in the
megaconstellation era remains sparse; the present paper is intended
in part to address that gap.

Complementary recent work has focused on \emph{extending} SGP4 to
close the gap with numerical propagators rather than measuring the
gap. SGP4-XP \citep{payne_sgp4xp_2022} extends the standard SGP4
analytic perturbation series with refined lunisolar gravity,
generalised resonance handling, solar radiation pressure for
high area-to-mass-ratio objects, and improved atmospheric density
modelling near Earth, while preserving the SGP4 evaluation budget
and the TLE input format. The differentiable-SGP4 line of work
\citep{acciarini_dsgp4_2025} reimplements the SGP4 analytic kernel
in a differentiable framework so the propagator can be fine-tuned
with gradient-based methods against observed ephemerides, and
composed with neural-network residual correctors. Our contribution
is orthogonal: we empirically characterise the baseline SGP4
versus high-fidelity behaviour on an operator megaconstellation
using only the public TLE format, and the per-cell
$(A,\,k)$ power-law table summarised in
Section~\ref{sec:results} is in a form usable as a benchmark
target by enhanced-propagator work of either flavour.

Operational LEO satellites maneuver, and a TLE pair whose interval
encloses a station-keeping burn violates the no-maneuver assumption
behind the next-TLE-as-truth comparison. \citet{lemmens2014}
introduced the standard public-domain technique for detecting
maneuvers from TLE histories: a jump in the semi-major axis derived
from the TLE mean motion, exceeding a per-platform threshold tuned
against the underlying orbit-determination noise floor, flags a
station-keeping event. Starlink's krypton Hall-effect station-keeping
cadence consists of continuous low-thrust burns rather than the
impulsive chemical maneuvers studied at the time of
\citet{lemmens2014}, so the threshold used here is calibrated
empirically against the bimodal $|\Delta a|$ distribution of the full
per-satellite TLE history within the analysis window
(Appendix~\ref{app:maneuver-filter},
Figure~\ref{fig:maneuver-filter}).

\section{Data and methods}
\label{sec:methods}

\subsection{TLE corpus and observation window}
\label{sec:methods-corpus}

We work from a single one-month observation window, 2026-04-01 to
2026-05-01 UTC, pinned in the repository as
\texttt{src/static/window.json} so that re-runs and Zenodo deposits
remain bit-identical. The Two-Line Element archive for that window
was retrieved once from the Space-Track \texttt{gp\_history} endpoint
restricted to \texttt{OBJECT\_NAME}~$\sim$~\texttt{STARLINK},
yielding 1{,}054{,}157 elements across 10{,}363 distinct catalog
identifiers. The raw cache is held outside the public repository to
honour Space-Track's redistribution terms; the downstream sample on
which all reported results depend is committed as
\texttt{src/static/tles\_cache.parquet}.

Starlink's operational update cadence of approximately six TLEs per
satellite per day --- compared with the once-per-day cadence assumed
in the prior-era literature \citep{vallado2006,vallado2012} --- has
two practical consequences for corpus construction. First, the
typical interval between consecutive TLEs for a single satellite is
of order four hours, well short of the propagation horizons of
interest; constructing pairs from strict consecutive elements would
cluster the entire corpus at a single staleness point. Second, the
dense per-satellite history makes maneuver detection at the
semi-major-axis-jump level statistically robust, since on the order
of 180 TLE epochs per satellite are available within the
observation window. We exploit both features by sampling starting
TLEs at one per satellite-day and searching for matching end TLEs at
four discrete target offsets (Section~\ref{sec:methods-pairs}), and
by running maneuver detection over the complete per-satellite
history rather than only on the pair endpoints. The geometric
distribution of the sampled satellites in
$(a,\,e,\,i)$ space is shown in Figure~\ref{fig:constellation-map}.

\subsection{Stratified satellite sampling}
\label{sec:methods-sampling}

Stratified sampling is performed over three altitude shells
corresponding to the active Starlink operational layers within the
analysis window. The shells are defined by the geocentric altitude
derived from each satellite's history-median TLE-mean-motion
semi-major axis: 533--547~km (the 540 shell), 547--557~km (the 550
shell), and 557--573~km (the 560 shell). Use of the per-satellite
median, rather than a single-TLE altitude, avoids misclassifying
satellites whose TLE history straddles an orbit-raising or
station-keeping maneuver. From the satellites whose median altitude
falls inside one of the three shells we draw 167 without replacement
per shell, using a fixed pseudo-random seed (\texttt{20260401}) and
yielding 501 sampled satellites. The (shell~$\times$~generation)
population of the resulting sample is set by Starlink's deployment
history rather than by the sampling procedure: the 540 shell carries
no v2-mini coverage in the analysis window, while the 550 and 560
shells contain a mix of generations
(Figure~\ref{fig:constellation-map}, Table~\ref{tab:corpus}).

From each sampled satellite's per-window TLE history we draw one
starting TLE per satellite-day --- specifically, the earliest
available epoch at or after 00{:}00 UTC of each day in the
analysis window. The one-per-day subsampling has two motivations.
First, Starlink's $\sim 6\times$/day publication cadence would
otherwise multiply-count nearly-identical operator OD epochs into
a corpus of strongly within-sat correlated pairs, inflating
effective sample size without adding statistical leverage.
Second, choosing the earliest daily epoch deterministically (rather
than a random TLE within each day) makes the corpus bit-reproducible
without recourse to an additional seed beyond
\texttt{20260401} above. The starting-TLE count of $14{,}917$ across
the 501 sampled sats reflects $\lesssim 30$ days of starting epochs
per sat, less the small number of sat-days with no TLE on record
($\lesssim 1\%$ of expected sat-day cells).

\subsection{Pair construction and maneuver filter}
\label{sec:methods-pairs}

For each sampled starting TLE we search the satellite's own history
for the element whose epoch is nearest to $t_{i} + \Delta t$ within a
$\pm 2$~h tolerance, for each target offset
$\Delta t \in \{6,\,24,\,72,\,168\}$~h. We adopt 6~h as the shortest
target offset because it lies just beyond Starlink's typical
$\sim 4$~h operator-update interval --- the boundary between fresh
and stale from the consumer's perspective and the regime in which
downstream consumer tools most often operate. The remaining offsets
at 1, 3, and 7 days span the short-, medium-, and long-staleness
ranges over which the power-law staleness curve assumed in
Hypothesis~H1 is statistically meaningful.

The $\pm 2$~h matching tolerance can in principle bias the corpus
against periods of poor operator tracking: a starting epoch
straddled by a multi-hour gap in the per-sat history would fail
to produce a pair at one or more $\Delta t$ targets. Across the
501-sat corpus and its $\sim 56{,}000$ in-window TLEs the median
inter-TLE interval is $4.8$~h (90th percentile $15.9$~h, 99th
percentile $23.9$~h, single maximum $108$~h); $17.6\%$ of
consecutive intervals exceed $12$~h, and the median satellite has
a longest within-window gap of $25.5$~h, rising to $41.7$~h at
the 95th percentile across sats. Empirically, $51\%$
of $(\text{starting~TLE} \times \Delta t)$ attempts produce a
matched pair within the $\pm 2$~h tolerance and the rest do not
($60.6\%$ unmatched at $\Delta t = 6$~h, $32.1\%$ at $1$~d,
$43.9\%$ at $3$~d, $59.3\%$ at $7$~d). The high unmatched
fractions are dominated by the Poisson-window character of the
matching --- the $4$~h target window is the same order as the
typical inter-TLE spacing, so a randomly-placed target window has
only an $\sim 50\%$ chance of containing any TLE even in dense
tracking conditions --- rather than by tracking-gap selection.
Of the $30{,}451$ matched candidates, $24{,}641$ ($81\%$) survive
the Section~\ref{sec:methods-pairs} maneuver filter to enter the
locked corpus.
The bias toward well-tracked sat-days is quantitatively bounded:
the 95th-percentile worst per-sat gap of $\sim 42$~h is well
inside the longest $\Delta t$ target ($168$~h), so no satellite is
systematically excluded; the conclusions of
Section~\ref{sec:results} characterise propagator behaviour during
sat-days the operator successfully covered. Both distributions are
visualised in
Figure~\ref{fig:selection-effect}~(Appendix~\ref{app:selection-effect}).

A pair $(\mathrm{TLE}_{i},\,\mathrm{TLE}_{j})$ is retained only if
no maneuver was detected anywhere inside the half-open interval
$(t_{i},\,t_{j}]$. The maneuver detector applies the
semi-major-axis-jump technique of \citet{lemmens2014} to every
consecutive TLE pair in the satellite's full history: the
consecutive change in the mean-motion-derived semi-major axis
$|\Delta a|$ is compared against a 100~m threshold, and any TLE
whose preceding step exceeds the threshold is flagged as the
timestamp of a station-keeping event. The threshold is calibrated
empirically against the $|\Delta a|$ histogram of the entire
Starlink TLE archive within the window
(Appendix~\ref{app:maneuver-filter},
Figure~\ref{fig:maneuver-filter}). The per-step formulation is the
dominant detection signature for Starlink's krypton Hall-effect
thrusters because the station-keeping cadence is burst-scheduled with
multi-hour gaps rather than continuous; for a hypothetical
steady-state low-thrust drift mode whose per-step $|\Delta a|$ never
crosses 100~m but which integrates to a meaningful displacement over
several days, the filter is in principle blind, a known limitation of
the SMA-jump approach formalised by \citet[\S 3.2]{lemmens2014}.
A threshold sensitivity at $\{50,\,100,\,200\}$~m on the same locked
corpus quantifies the effect of that limitation on the headline
medians and is reported in
Appendix~\ref{app:maneuver-threshold-sensitivity}.
After applying the filter, the
locked sweep corpus contains 24{,}641 surviving pairs; the
decomposition by altitude shell and Starlink generation is given in
Table~\ref{tab:corpus}.

\begin{table}[ht]
  \centering
  \caption{Composition of the locked sweep corpus by altitude shell
  and Starlink generation. Counts give the number of surviving
  $(\mathrm{TLE}_{i},\,\mathrm{TLE}_{j})$ pairs after the maneuver
  filter; column and row totals aggregate across the four
  $\Delta t$ buckets.}
  \label{tab:corpus}
  \begin{tabular}{lrrrr}
    \toprule
    Shell (km) & v1.0 & v1.5 & v2-mini & Total \\
    \midrule
    540   &  52 & 7{,}427 & ---     & 7{,}479 \\
    550   &  37 & 2{,}586 & 6{,}908 & 9{,}531 \\
    560   &  65 & 4{,}629 & 2{,}937 & 7{,}631 \\
    \midrule
    Total & 154 & 14{,}642 & 9{,}845 & 24{,}641 \\
    \bottomrule
  \end{tabular}
\end{table}

\subsection{Spacecraft properties}
\label{sec:methods-spacecraft-props}

Per-satellite dry mass is taken from McDowell's General Catalog of
Artificial Space Objects \citep{mcdowell2020}, which resolves at the
launch-batch level across the Starlink generations in our corpus.
Generation (v1.0 / v1.5 / v2-mini) is classified from the same catalog's
\texttt{PLName} field.

\subsection{Attitude regime and drag-area choice}
\label{sec:methods-attitude}

A Starlink satellite at its operational altitude maintains a
``shark-fin'' attitude --- solar arrays held edge-on to the ram
direction and phased-array antennas pointing toward nadir --- as the
station-keeping configuration. A high-drag ``open-book'' attitude is
engaged only during autonomous collision-avoidance maneuvers and
controlled de-orbit. SpaceX's most recent semi-annual constellation
status report \citep{spacex_fcc_semiannual_2024} attributes an average
of $14$ thruster firings per satellite over the six-month window
1~December 2023 through 31~May 2024 to autonomous collision
avoidance, against continuous on-station thrust for station-keeping
\citep{lang2025}; the open-book residence fraction is therefore small,
event-driven, and concentrated on intervals that the maneuver filter
(Section~\ref{sec:methods-pairs}, Appendix~\ref{app:maneuver-filter})
removes from the corpus before any pair reaches the sweep.

\begin{table}[ht]
  \centering
  \caption{Per-attitude effective drag areas at the two $C_D$
  conventions, with the per-generation values used in the main sweep
  in the bottom block. Baruah ram areas are quoted at $C_D = 1$; the
  sweep uses the free-molecular-flow $C_D = 2.2$ convention, so the
  effective area at $C_D = 2.2$ is the Baruah value divided by $2.2$.
  The v2-mini row has no published ballistic-coefficient analysis and
  is scaled from the v1.x shark-fin value by the bus-size ratio
  reported in \citet{krebs_starlink_v2mini}.}
  \label{tab:drag-areas}
  \begin{tabular}{lrrr}
    \toprule
                                       & $C_D\!=\!1$ & $C_D\!=\!2.2$ & Ratio vs.\\
    Attitude / configuration           & area (m$^2$) & area (m$^2$) & open-book \\
    \midrule
    Open-book (CAM / de-orbit)\,\citep{baruah2024} & 1.00 & 0.45 & 1.00$\times$ \\
    Shark-fin (operational)\,\citep{baruah2024}    & 4.48 & 2.04 & 4.48$\times$ \\
    \midrule
    v1.0 / v1.5 (sweep, shark-fin)                 & 4.4  & 2.0  & 4.40$\times$ \\
    v2-mini (sweep, shark-fin $\times$ bus-size)   & 9.9  & 4.5  & 9.90$\times$ \\
    \bottomrule
  \end{tabular}
\end{table}

We adopt the shark-fin value as the duty-cycle-representative drag
area: $2.0\,$m$^2$ at $C_D = 2.2$ for v1.0 and v1.5 (matching
Baruah's $4.48\,$m$^2$ at $C_D = 1$ to within the rounding in
Table~\ref{tab:drag-areas}), and $4.5\,$m$^2$ at $C_D = 2.2$ for
v2-mini (scaled from the v1.x shark-fin value by a factor of $2.25$
to reflect the larger v2-mini bus). The full attitude swing from
shark-fin to open-book is a factor of $4.48\times$ change in
$C_D \cdot A$, which is well outside the $\pm 20\%$ uncertainty range
explored in Appendix~\ref{app:cda-sensitivity}; the sensitivity test
is calibrated to bound the bus-scaling and mass-estimate uncertainty
within the operational attitude, not to span the attitude swing
itself (which is excluded from the corpus by construction).

\subsection{Force model and integrator}
\label{sec:methods-forcemodel}

The high-fidelity arm of the comparison propagates each starting state
forward to $t_{j}$ using NASA's General Mission Analysis Tool (GMAT)
release R2026a, configured as in Table~\ref{tab:force-model}. The
static configuration is identical for every run; the per-pair
quantities (epoch, Cartesian state in MJ2000Eq, dry mass, drag and
solar-radiation-pressure coefficients and effective areas, and the
integration duration) are injected into the mission script through
GMAT's field-override interface ahead of each call to the mission
sequence.

\begin{table}[ht]
  \centering
  \caption{High-fidelity force-model and integrator configuration
  applied to every sweep run. The values listed are identical across
  all 24{,}641 propagations; per-pair quantities (initial state,
  epoch, dry mass, drag and SRP coefficients and effective areas,
  propagation duration) are supplied through GMAT's field-override
  interface and are not shown here.}
  \label{tab:force-model}
  \begin{tabular}{lll}
    \toprule
    Component & Setting & Reference \\
    \midrule
    Central body                 & Earth                              & --- \\
    Gravity field                & EGM2008, degree~$70$, order~$70$   & \citet{pavlis2012} \\
    Third bodies                 & Sun, Moon (point masses)           & --- \\
    Solar radiation pressure     & Enabled, conical Earth shadow      & GMAT R2026a default \\
    Atmospheric drag             & NRLMSISE-00                        & \citet{picone2002} \\
    \midrule
    Integrator                   & Runge--Kutta $8(9)$                & \citet{verner1978} \\
    Accuracy tolerance           & $10^{-12}$~km                       & --- \\
    Initial step                 & $60$~s                              & --- \\
    Minimum / maximum step       & $10^{-3}$~s / $600$~s               & --- \\
    Error-control mode           & RSS step                            & --- \\
    \bottomrule
  \end{tabular}
\end{table}

EGM2008 is loaded at degree and order $70$ from a project-installed
coefficient file produced by parsing the NGA flat-text
\texttt{EGM2008\_to2190\_TideFree} distribution into the GMAT
\texttt{.cof} format; the truncation error at degree $70$ relative
to the full $2160\!\times\!2160$ expansion is small enough at LEO
altitudes \citep{pavlis2012} that it sits well below the integrated
drag and operator-OD-residual contributions to the kilometre-scale
position errors characterised by this analysis, and is not
load-bearing for any of the comparisons that follow. Atmospheric drag uses GMAT's NRLMSISE-00
implementation \citep{picone2002}, the SSA community standard for
thermospheric density at LEO altitudes; the integrator reads CSSI-format
F10.7 and Ap inputs from the project-local space-weather file described
in Section~\ref{sec:methods-reproducibility}. The F10.7 and Ap values
reported per output row in Section~\ref{sec:methods-metrics} are an
independent analysis annotation joined onto each row at
post-processing time and play no role in the integration. The $8(9)$-order Runge--Kutta scheme
with embedded error estimator \citep{verner1978} held the configured
tolerance across the longest seven-day arcs in the corpus with no
observed integration failures (manifest tally: 24{,}641
ok\,/\,$0$\,failed).

The Earth-gravity-parameter convention differs between the
SGP4/TLE stack and the EGM2008 force model: SGP4 uses WGS-72
($\mu = 398{,}600.8$~km$^3$/s$^2$, $R_{\oplus} = 6378.135$~km),
whereas EGM2008 uses the more recent WGS-84 values
($\mu = 398{,}600.4418$~km$^3$/s$^2$, $R_{\oplus} = 6378.137$~km).
The fractional offset $\delta\mu / \mu \approx 1\times 10^{-6}$
aliases into a semi-major-axis bias of similar order
($\mathcal{O}(10)$~m at LEO) when the SGP4-derived state is
re-interpreted as an EGM2008-osculating initial condition --- not
load-bearing for the kilometre-scale errors characterised in this
paper, but cited explicitly here so the constant mismatch does
not look unacknowledged.

\subsection{Frame and time conventions}
\label{sec:methods-frame-time}

TEME (True Equator, Mean Equinox of date) is the pseudo-inertial
frame in which SGP4 evaluates its mean-element ephemeris. Unlike
J2000 or ICRF, TEME's polar axis is the true equator of date and
its origin in right ascension is the \emph{mean} equinox; the
resulting frame is approximately equivalent to J2000 ECI for short
propagation arcs but accumulates a few milliarcseconds per year of
drift due to nutation neglect in the right-ascension reference.
Two conventions of TEME appear in the literature --- the modern
Vallado formulation, used here and implemented in the IAU SOFA /
\texttt{erfa} stack, and an older STK ``true-of-date'' variant
\citep[\S\,3.7]{vallado2013}; we adopt the Vallado/erfa convention,
which is the community-canonical choice for open-source TLE
tooling. GMAT's MJ2000Eq frame is the J2000 mean-equator /
mean-equinox inertial frame --- equivalent to ``J2000 ECI'' in
non-GMAT vocabulary --- and is the inertial frame the
high-fidelity propagator integrates in.

Initial states for the GMAT propagation are derived from the starting
TLE by evaluating SGP4 at $\Delta t = 0$ in TEME and rotating into
MJ2000Eq. The TEME-to-MJ2000Eq rotation follows
Algorithm 24 of \citet{vallado2013}, composed from three primitives
in the IAU SOFA / \texttt{erfa} library: the IAU 1976 precession
matrix (\texttt{pmat76}), the IAU 1980 nutation matrix
(\texttt{nut80} together with \texttt{numat}), and the 1994 equation
of equinoxes (\texttt{eqeq94}). The underlying $\mathrm{UTC}\to
\mathrm{TT}$ transform is supplied by an IERS-aware time module
referencing the \citet{iers2010} conventions. The same rotation
matrix is applied to the velocity vector; the residual error from
neglecting $\mathrm{d}\mathbf{M}/\mathrm{d}t$ in the velocity rotation
is bounded above by the precession rate
($\lesssim 50{}^{\prime\prime}\,\mathrm{yr}^{-1}$) and the peak
nutation amplitude ($\lesssim 9{}^{\prime\prime}$), translating to
$\lesssim 10^{-5}$~mm
s$^{-1}$ at LEO orbital speeds --- several orders of magnitude below
the kilometre-scale position errors characterised in this paper.
SGP4 itself is evaluated through the \texttt{sgp4} Python
implementation of \citet{rhodes_sgp4}, the canonical port of the
Vallado open-source reference code, which uses the WGS-72 gravity
constants that the TLE format was constructed against.

\subsubsection{Dynamical-consistency caveat}
\label{sec:methods-dynamical-consistency}

The Cartesian state delivered by $\mathrm{SGP4}(\mathrm{TLE},\,
\Delta t = 0)$ is \emph{not} an osculating Cartesian state in the
sense an external high-fidelity propagator expects. SGP4 is an
analytic averaged-element propagator built on the Brouwer
mean-element formulation \citep[\S\,11.6]{vallado2013}: a TLE
encodes mean Keplerian elements at epoch, and SGP4's analytic
kernel \citep{hoots2004} unwraps those mean elements through
short-period perturbation series to produce a Cartesian
position--velocity pair. That pair is a pseudo-osculating
intermediate, not the instantaneously osculating state of the
underlying physical orbit. Numerically integrating it as if it
were osculating --- as we do when feeding it to GMAT --- carries
the SGP4 short-period kernel's intrinsic amplitude into the
initial condition, bounded below by the short-period perturbation
scale of order $\mathcal{O}(1)$~km at LEO. The mismatch is
fundamental to the comparison this paper performs: the
high-fidelity arm inherits an initial condition that lives in a
different dynamical reference than the truth construction (which
re-evaluates SGP4 at the matched ending TLE), so part of the
hi-fid-vs-SGP4 gap reported in Section~\ref{sec:results} is the
cost of leaving the SGP4 mean-element manifold rather than a
deficiency of the integrator. This is one of the three mechanisms
behind the H2 negative result discussed in
Section~\ref{sec:discussion}.

\subsection{Error metrics and solar context}
\label{sec:methods-metrics}

For each sweep run we compute two position-error vectors at the end
epoch $t_{j}$, both in MJ2000Eq. The first,
$\Delta\mathbf{r}_{\mathrm{SGP4}}$, is the difference between the
state obtained by propagating the starting TLE with SGP4 to $t_{j}$
and the proxy truth (next-TLE state) at $t_{j}$, constructed by
evaluating SGP4 at $\Delta t = 0$ on the matched ending TLE. The
second, $\Delta\mathbf{r}_{\mathrm{hifi}}$, is the corresponding
difference for the GMAT high-fidelity propagation initialised from
the same starting state in MJ2000Eq. The L$_{2}$ norms
$\lVert\Delta\mathbf{r}_{\mathrm{SGP4}}\rVert$ and
$\lVert\Delta\mathbf{r}_{\mathrm{hifi}}\rVert$ drive the staleness
curves of Figures~\ref{fig:sgp4-growth} and~\ref{fig:hifi-growth},
the per-pair paired-scatter of
Figure~\ref{fig:propagator-scatter}, and the per-cell power-law
fits of Figure~\ref{fig:powerlaw-fits} and Table~\ref{tab:powerlaw}.

We additionally decompose each error vector into radial, along-track,
and cross-track components in the local orbital frame of the
next-TLE proxy at $t_{j}$. The basis follows the standard RSW construction
\citep[\S\,3.4]{vallado2013}:
$\hat{\mathbf{e}}_{r} = \hat{\mathbf{r}}_{\mathrm{truth}}$;
$\hat{\mathbf{e}}_{h} = \widehat{(\mathbf{r}\times\mathbf{v})}
_{\mathrm{truth}}$, along the angular-momentum direction
(cross-track); and
$\hat{\mathbf{e}}_{t} = \hat{\mathbf{e}}_{h}\times\hat{\mathbf{e}}_{r}$,
completing the right-handed triad (along-track; approximately the
velocity direction for the near-circular orbits in our corpus). The
decomposition separates phase-only mistrack along the velocity
direction from genuine geometric error in the plane perpendicular to
it, and is the natural frame in which to compare propagator skill
component by component; results are reported in
Figure~\ref{fig:error-decomposition}.

Each output row also carries the daily observed F10.7 solar flux
and the planetary daily Ap index for the UTC date of the starting-TLE
epoch, parsed from CelesTrak's \texttt{sw19571001.txt} space-weather
archive and cached locally in \texttt{src/static/sw\_cache.parquet}.
These two columns are post-processing analysis annotations used only
by the H3 regression of the per-satellite SGP4 staleness coefficient
against solar activity (Figure~\ref{fig:solar-modulation}); they
play no role in the integration itself, which reads its own
space-weather inputs through GMAT's bundled NRLMSISE-00
implementation (Section~\ref{sec:methods-forcemodel}).

The CSSI space-weather inputs consumed by GMAT's NRLMSISE-00 driver
are read from a project-local snapshot of CelesTrak's
\texttt{sw19571001.txt} (Section~\ref{sec:methods-reproducibility}),
so the hi-fid integrator sees the same daily-observed $F_{10.7}$ and
$Ap$ values over the analysis window as the per-row annotation
described above.

\subsubsection{Statistical estimators}
\label{sec:methods-statistics}

The per-cell power-law fit reported in
Figure~\ref{fig:powerlaw-fits} and Table~\ref{tab:powerlaw} is a
weighted-with-unit-weights ordinary-least-squares regression of
$\log_{10}\lVert\Delta\mathbf{r}\rVert$ on
$\log_{10}(\Delta t / 1\,\text{h})$, fit on every pair in the
(altitude shell $\times$ pooled generation $\times$ propagator) cell.

Each pair contributes its own measured $\Delta t$ (the
\texttt{actual\_dt\_sec} column of the per-run parquet, which carries
the realised epoch spacing inside the $\pm 2$~h matching tolerance);
the four nominal staleness buckets $\{6\,\text{h},\,1\,\text{d},\,
3\,\text{d},\,7\,\text{d}\}$ used by
Figures~\ref{fig:sgp4-growth}, \ref{fig:hifi-growth}, and
\ref{fig:error-decomposition} are a visualisation device only and do
not enter the fit. A well-populated cell carries
$\mathcal{O}(10^{3})$ pairs, so the regression has $\mathcal{O}(10^{3})$
degrees of freedom, not the four that a naive reading of the
bucketed figures would suggest; per-pair fitting also lets the
within-bucket spread in $\Delta t$ contribute information that a fit
to per-bucket medians would discard.

Per-pair weights are uniform: pair-level
$\mathrm{SE}(\log_{10}\lVert\Delta\mathbf{r}\rVert)$ is
not available on the run parquet, and the satellite-level bootstrap
introduced below absorbs the population scatter that an
inverse-variance scheme would otherwise track.

Confidence intervals for $A$ and $k$ are percentile intervals from a
$1{,}000$-resample bootstrap drawn at the satellite level: each
resample picks satellites with replacement from the cell and pools
all four bucket-of-pairs from each drawn satellite. Resampling at
the satellite rather than the pair level preserves within-satellite
correlation across the four staleness buckets --- pairs sharing a
ballistic coefficient, attitude duty cycle, and perturbation history
are not independent observations, and a pair-level bootstrap would
understate uncertainty by treating them as such. Reported CIs are
correspondingly tight on well-populated cells because $n \approx 167$
satellites per shell sample a geometrically narrow $(a, e, i)$
neighbourhood, not because the resample under-accounts for the
within-satellite covariance the construction specifically targets;
the mixed-effects cross-check in Appendix~\ref{app:mixed-effects}
verifies that the parametric slope agreement holds across cells. Asymmetric
percentile intervals are reported as such; bias-corrected and
accelerated (BCa) intervals are not adopted as a default because the
percentile intervals are visually symmetric in every populated cell,
but the same satellite-level resampling is BCa-ready and could be
swapped if a future re-run developed a skewed estimator.

Table~\ref{tab:powerlaw} additionally reports the per-cell
coefficient of determination $R^{2}$ of the unconstrained
per-pair fit and likelihood-ratio test (LRT) p-values for the two
physically interpretable nulls $k = 1$ (constant mean-motion bias,
linear along-track displacement) and $k = 2$ (constant unmodelled
along-track acceleration, quadratic along-track displacement). The
LRT statistic for the slope constraint $k = k_{0}$ is the
asymptotically $\chi^{2}_{1}$ quantity
$n\,\log(\mathrm{RSS}_{c} / \mathrm{RSS}_{u})$, where
$\mathrm{RSS}_{u}$ is the residual sum of squares of the
unconstrained fit and $\mathrm{RSS}_{c}$ is the analogous quantity
under $k = k_{0}$ with the intercept profiled out by recentering on
$\overline{y - k_{0} x}$. The LRT p-value is the survival probability
of the test statistic under that asymptotic distribution. We do
\emph{not} report bootstrap CIs for $R^{2}$ or for the LRT
p-values: both are reported as point estimates from the full-cell
unconstrained fit, since they answer questions about the
goodness-of-fit and the discriminability of the slope nulls rather
than about the fitted parameter values themselves.

Appendix~\ref{app:mixed-effects} reports a parametric cross-check
of the main estimator: a per-cell linear mixed-effects fit of
$\log_{10}\lVert\Delta\mathbf{r}\rVert$ on $\log_{10}\Delta t$
with a satellite-level random intercept,
$\log_{10}\lVert\Delta\mathbf{r}\rVert \sim
\log_{10}\Delta t + (1 \mid \texttt{norad\_id})$.
Mixed effects treat the same within-satellite correlation
parametrically, by assuming a normal random-intercept distribution
across satellites; the bootstrap handles it non-parametrically. A
substantive disagreement between the two on any cell would be a
signal to escalate the mixed-effects fit to the primary estimator
in a follow-up revision. The per-cell comparison and a
mixed-effects-versus-bootstrap agreement summary are deferred to
the appendix so the main-body table remains the single estimator
of record.

\subsubsection{Physical interpretation of the power-law exponent}
\label{sec:methods-powerlaw-physics}

The power-law form
$\lVert\Delta\mathbf{r}(\Delta t)\rVert \approx A\,\Delta t^{k}$
is descriptive rather than first-principles, but its two
physically interpretable limits $k = 1$ and $k = 2$ that the LRT
in Section~\ref{sec:methods-statistics} tests against correspond
to specific error-source regimes in the orbit-propagation budget,
both expressed through the along-track component on which the
gross error overwhelmingly lives
(Section~\ref{sec:results} and Figure~\ref{fig:error-decomposition}).

Consider first a constant mean-motion error $\delta n$ at epoch ---
the operational footprint of a $B^{\!\star}$ mis-fit, where the
operator-fitted ballistic coefficient at $t_{i}$ encodes a mean
drag environment over the OD window that differs systematically
from the conditions encountered during forward propagation.
Integrating the rotational equation $\dot{\theta} = n + \delta n$
over $(t_{i},\,t_{j}]$ gives an along-track angle offset
$\delta\theta = \delta n\,\Delta t$, which maps to an arc-length
displacement
\begin{equation}
  \label{eq:powerlaw-k1}
  \Delta s(\Delta t) \;\approx\; a\,\delta n\,\Delta t,
\end{equation}
i.\,e.\ $k = 1$ with $A = a\,\delta n$ (in suitable units for the
$\Delta t = 1\,\text{h}$ normalisation Table~\ref{tab:powerlaw}
adopts). The displacement is linear in $\Delta t$ because the
mean-motion error is constant; no integration of an acceleration
is involved.

Consider now a constant unmodelled in-track acceleration $a_{\rm{res}}$
--- the operational footprint of a drag-area or
spacecraft-property mis-specification, where the integrator (or,
implicitly, SGP4's own averaged drag term) sees a $\rho\,C_{D}\,A$
product offset from the truth by a constant. Integrating
$\ddot{s} = a_{\rm{res}}$ twice with zero displacement and velocity
at $t_{i}$ gives
\begin{equation}
  \label{eq:powerlaw-k2}
  \Delta s(\Delta t) \;\approx\; \tfrac{1}{2}\,a_{\rm{res}}\,\Delta t^{2},
\end{equation}
i.\,e.\ $k = 2$ with $A = a_{\rm{res}} / 2$ (again in
$\Delta t = 1\,\text{h}$ units).

The empirical $k$ values across our populated cells span both sides
of unity --- Table~\ref{tab:powerlaw} reports $k$ between
$0.75$ and $1.44$ across the SGP4 cells and $k$ between $0.99$ and
$1.28$ across the high-fidelity cells --- which is the expected
signature of a \emph{mixture} of the two error modes operating
simultaneously: a partial $B^{\!\star}$ mis-fit at epoch dragging
$k$ toward unity (Eq.~\eqref{eq:powerlaw-k1}) and a residual
constant-acceleration term dragging it toward two
(Eq.~\eqref{eq:powerlaw-k2}), with the cohort-specific balance
between them setting the per-cell exponent. The per-cell $(A,\,k)$ pair is, on this
reading, a coarse two-parameter summary of the operator's
epoch-time error budget for that satellite cohort: $A$ scales
with the joint magnitude of both contributions evaluated at
$\Delta t = 1\,\text{h}$, and $k$ encodes the mix --- closer to
unity when mean-motion bias dominates, closer to two when a
residual constant in-track acceleration dominates. Other
contributions (initial-state error mapped forward by the state
transition matrix, time-varying drag through a non-constant
$\rho$, gravity-truncation residuals) shift the effective $(A,
\,k)$ pair without changing the two limiting endpoints; the LRT
in Section~\ref{sec:methods-statistics} discriminates a cell
against those endpoints with the same $\chi^{2}_{1}$ statistic
in both directions.

\subsubsection{H3 regression specification}
\label{sec:methods-h3-regression}

Hypothesis H3 asks whether the per-satellite SGP4 staleness
coefficient $A_{i}$ tracks the solar-EUV proxy F10.7 in the
drag-dominant cohorts of the corpus. The model choice follows
from a short derivation. Atmospheric drag at LEO is
$\mathbf{a}_{\mathrm{drag}} = -\tfrac{1}{2}\,\rho(h,
\mathrm{F}10.7)\,v^{2}\,(C_{D} A / m)\,\hat{\mathbf{v}}_{\mathrm{rel}}$;
SGP4 absorbs the time-averaged $\tfrac{1}{2}(C_{D} A / m)\rho_{\mathrm{ref}}$
encountered during the OD window into a single fitted scalar
$B^{\!\star}$, so the operator's epoch-time fit lags actual
density whenever F10.7 trends within the prediction span. The
forward-propagation along-track error from that lag is
proportional to $\Delta\rho(\mathrm{F}10.7)/\rho_{\mathrm{ref}}$
integrated over $\Delta t$. Taking logs of the underlying
multiplicative chain,
\begin{equation}
  \label{eq:h3-derivation}
  \log_{10} A_{i}
  \;\approx\;
  \log_{10}(C_{D} A / m)_{i}
  \;+\;
  \log_{10}\rho_{\mathrm{ref}}\!\bigl(h_{i},\,\mathrm{F}10.7_{i}\bigr)
  \;+\;
  \mathrm{const},
\end{equation}
and differentiating in F10.7 at fixed altitude isolates the
quantity H3 is asking about:
\begin{equation}
  \label{eq:h3-slope-altitude-only}
  \frac{\partial\,\log_{10} A}{\partial\,\mathrm{F}10.7}
  \,\Big|_{h}
  \;=\;
  \frac{\partial\,\log_{10}\rho}{\partial\,\mathrm{F}10.7}
  \,\Big|_{h}.
\end{equation}
Spacecraft properties enter the intercept of
Eq.~\eqref{eq:h3-derivation} through the per-satellite
$(C_{D} A / m)$ term, but they drop out of
Eq.~\eqref{eq:h3-slope-altitude-only} entirely. The H3 slope is
therefore predicted to be a function of altitude and the local
density-modulation kernel only, and not of platform generation.

We accordingly fit one slope per altitude shell, with the
intercept differences between pooled generations absorbed as a
nuisance covariate. The model is the additive ANCOVA form
\begin{equation}
  \label{eq:h3-model}
  \log_{10} A_{i}
  \;=\;
  \alpha_{\mathrm{gen}(i)}
  \;+\;
  \beta\,\mathrm{F}10.7_{i}
  \;+\;
  \varepsilon_{i},
\end{equation}
fit by ordinary least squares per shell. The
$A_{i}$ entering Eq.~\eqref{eq:h3-model} is the per-satellite
SGP4 power-law coefficient recovered by the
Section~\ref{sec:methods-statistics} fit restricted to a single
satellite. The fit is restricted to the SGP4 arm because the
operator-fitted $B^{\!\star}$ is the quantity
Eq.~\eqref{eq:h3-slope-altitude-only} concerns; a parallel hi-fid
regression on $A_{i,\mathrm{hifi}}$ is a natural follow-up.

The naive alternative --- a per-shell pooled fit that drops the
$\alpha_{\mathrm{gen}}$ term --- fails empirically at $550$ and
$560$~km. The v1.x and v2-mini cohorts share an altitude shell
but carry per-satellite $A$ medians that differ by roughly a
factor of four (v1.x at $0.13$--$0.22$, v2-mini at $0.06$~km at
$\Delta t = 1$~h), reflecting the larger $C_{D} A / m$ scale of
the newer bus together with operator OD differences across the
two generations. Their epoch distributions also sample slightly
different mean F10.7 within the analysis window. A pooled
$550$~km regression that ignores the intercept gap aliases that
cross-cohort offset into a wrong-signed slope:
$\hat\beta = -0.054$~per~sfu at $p \approx 2\times 10^{-7}$,
diagnostic-run on the same corpus. The ANCOVA form of
Eq.~\eqref{eq:h3-model} absorbs the offset into
$\alpha_{\mathrm{gen}}$ and exposes the within-cohort slope that
Eq.~\eqref{eq:h3-slope-altitude-only} actually predicts.

For each shell we additionally fit the interaction-augmented
model
$\log_{10} A_{i} = \alpha_{\mathrm{gen}(i)} +
(\beta + \delta\beta_{\mathrm{gen}(i)})\,\mathrm{F}10.7_{i} +
\varepsilon_{i}$
and report the partial-$F$ test of
$H_{0}\!: \delta\beta_{\mathrm{gen}} = 0$ against the additive
model. Under Eq.~\eqref{eq:h3-slope-altitude-only} the additive
restriction is the physics expectation; rejection of $H_{0}$ is
informative as a sign that the simple drag-mismodelling picture
omits a cohort-specific F10.7 response we have not modelled, not
as a license to substitute the interaction model as the headline
estimator.

Two predictors are reported per shell. The headline predictor
is the per-satellite mean of the daily observed F10.7 over the
satellite's window of starting epochs, parsed from the CelesTrak
cache described in Section~\ref{sec:methods-metrics}. As a
robustness check we additionally fit Eq.~\eqref{eq:h3-model}
with the per-satellite mean of the CelesTrak 81-day-centred
F10.7 average as the predictor; the 81-day average is the
long-term thermospheric driver that NRLMSISE-00 consumes
internally and is the convention-matched predictor for any
attempt to interpret $\hat\beta$ as a density-modulation slope.

Over the 30-day corpus window the daily-mean predictor spans
roughly $17$~sfu sat-to-sat ($\approx 104$--$138$), while the
81-day-centred predictor compresses to $\approx 1.3$~sfu
($\approx 124.6$--$125.9$) by construction. The 81-day
robustness fit has very limited statistical leverage on the
present window, so we treat the daily-mean fit as the
load-bearing report.

For each (shell, predictor) cell the full-sample OLS of
Eq.~\eqref{eq:h3-model} gives the point estimates
$(\hat\alpha_{\mathrm{gen}},\,\hat\beta)$ with asymptotic
standard errors $\mathrm{SE}(\hat\beta)$ and the two-sided
$t$-stat $p$-value for $H_{0}\!: \beta = 0$ under standard OLS
assumptions on the residuals. The $95\%$ percentile confidence
interval on $\hat\beta$ is taken from a $1{,}000$-resample
satellite-level bootstrap that draws sats with replacement from
the shell and refits the additive model; CI bounds are the
$2.5$th and $97.5$th percentiles of the resampled $\hat\beta$
distribution. The same bootstrap supplies per-gen intercept-offset
CIs through the resampled $\hat\alpha_{\mathrm{gen}}$ values, so
the figure's per-gen CI ribbons combine the slope and
intercept-offset uncertainties consistently. The resampling
structure is identical to that of the per-cell power-law CIs in
Section~\ref{sec:methods-statistics}.

Empirical results are reported in
Figure~\ref{fig:solar-modulation} and discussed in
Section~\ref{sec:discussion}. The daily-mean fit returns, per
shell:
\begin{itemize}
  \item $540$~km (v1.x only):
  $\hat\beta = +0.017$~per~sfu ($95\%$~CI $[-0.002,\,+0.036]$,
  $p = 0.07$, $R^{2} = 0.02$).
  \item $550$~km (v1.x, v2-mini):
  $\hat\beta = -0.009$~per~sfu ($95\%$~CI $[-0.029,\,+0.009]$,
  $p = 0.31$, $R^{2} = 0.45$), with the slope consistent with
  zero and the interaction test failing to reject the additive
  model ($p_{\beta\times\mathrm{gen}} = 0.69$).
  \item $560$~km (v1.x, v2-mini):
  $\hat\beta = +0.014$~per~sfu ($95\%$~CI
  $[+0.001,\,+0.031]$, $p = 0.034$, $R^{2} = 0.24$) --- the
  smallest positive slope to clear conventional significance on
  this corpus.
\end{itemize}
The high $R^{2}$ values at $550$ and $560$~km reflect the
intercept covariate absorbing the cross-cohort baseline gap, not
an explanatory contribution from F10.7. The interaction test at
$560$~km does reject the additive restriction
($p_{\beta\times\mathrm{gen}} = 0.005$); the narrow predictor
span and modest cell counts make the departure from
Eq.~\eqref{eq:h3-slope-altitude-only} difficult to attribute
cleanly to either a real cohort-specific F10.7 response or to a
window-end sampling pattern, and we flag it rather than re-fit.

The 81-day-centred robustness fits, refit on the same shells
with the same ANCOVA design, return
$\hat\beta = +0.003$~per~sfu ($95\%$~CI $[-0.296,\,+0.241]$,
$p = 0.98$) at $540$~km,
$\hat\beta = -0.432$~per~sfu ($95\%$~CI $[-0.703,\,-0.205]$,
$p = 0.001$) at $550$~km, and
$\hat\beta = -0.253$~per~sfu ($95\%$~CI $[-0.473,\,+0.008]$,
$p = 0.015$) at $560$~km. The negative slopes at $550$ and
$560$~km are physically wrong-signed --- the local
$\partial\,\log\rho / \partial\,\mathrm{F}10.7$ at LEO altitudes
is uniformly positive, so a genuine drag-modulation signal would
return $\hat\beta > 0$ --- and we read the apparent significance
as a sampling artefact rather than a competing H3 estimate. The
$1.3$~sfu pooled span of the 81-day-centred predictor compresses
within each cohort to $0.57$--$1.24$~sfu, narrow enough that each
satellite's per-sat mean is set largely by which sub-window of
April~2026 its starting epochs cluster in; any window-internal
correlation between epoch placement and per-sat $A$ then maps
into the slope rather than into the $\alpha_{\mathrm{gen}}$
covariate. We report the 81-day fits for completeness and
treat the daily-mean fit as the load-bearing H3 estimate on the
present 30-day corpus.

Section~\ref{sec:discussion}
returns to whether the absence of a clear $|\hat\beta_{540}|
\gtrsim |\hat\beta_{560}|$ altitude-attenuation ordering reflects
a genuine cohort-level cancellation or the narrowness of the
F10.7 lever available in the analysis window.

\subsection{Reproducibility surface}
\label{sec:methods-reproducibility}

The source repository contains the complete data-preparation
pipeline (TLE fetch, sampling, pair construction, maneuver filter,
spacecraft-property attachment), the GMAT mission script, the
parallel sweep driver, the aggregation and statistics utilities,
every figure-generation script, and the manuscript \LaTeX{} source,
all released under the MIT licence at the URL given in
Section~\ref{sec:availability}. The sweep output bundle --- per-run
parquets, the resumable manifest, and the mission script used to
generate them --- is deposited on Zenodo at the version DOI listed
in the build configuration; the manuscript build target fetches that
bundle and rebuilds every figure from the cached data, so
re-rendering the PDF from a clean checkout does not require a local
GMAT installation. Re-running the full sweep from scratch does
require GMAT R2026a, the EGM2008 coefficient file installed through
the project's installer script, and Space-Track credentials for the
one-time TLE fetch.

The CSSI-format space-weather text file consumed by GMAT's NRLMSISE-00
driver is committed as a frozen snapshot of CelesTrak's
\texttt{sw19571001.txt} in the repository under
\texttt{src/static/\allowbreak SpaceWeather-\allowbreak All-\allowbreak v1.2.\allowbreak txt}; it is regeneratable
from the CelesTrak source via \texttt{make fetch-gmat-sw}. The sweep
driver supplies the snapshot's absolute path to GMAT through the
\texttt{FM.Drag.\allowbreak CSSISpaceWeatherFile} script-level
override, so the integrator reads observed daily $F_{10.7}$ and $Ap$
values over the entire analysis window rather than the predicted
inputs that GMAT R2026a's shipped CSSI file would supply for epochs
past its observed-data horizon. The committed snapshot covers the
April 2026 corpus window with daily observed data through 2026-05-16.

Exact versions of the load-bearing
astrodynamics dependencies --- SOFA / \texttt{erfa} for the
TEME-to-MJ2000Eq rotation, the \texttt{sgp4} Python implementation
of \citet{rhodes_sgp4} for the SGP4 evaluation, and \texttt{astropy}
for IERS-aware time conversion --- are pinned in
\texttt{pyproject.toml} and the conda environment lockfile shipped
in the repository, so the rotation chain and SGP4 numerical kernel
are bit-identical across re-runs.

\section{Results}
\label{sec:results}

\begin{figure}[ht]
  \centering
  \includegraphics[width=\linewidth]{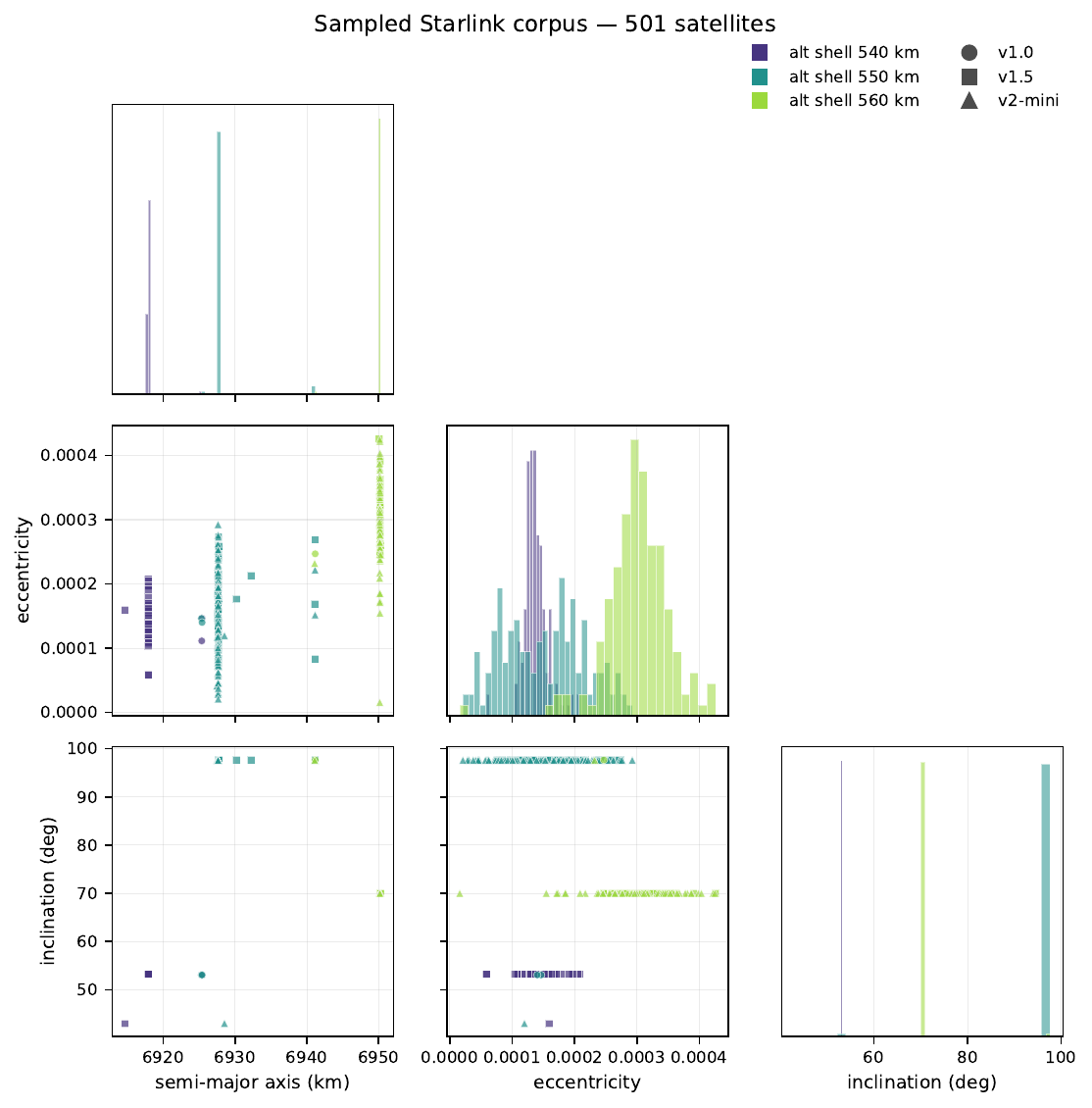}
  \caption{Sampled corpus. Pairwise scatter of semi-major axis,
  eccentricity, and inclination for the $501$ Starlink satellites in
  the analysis sample. Altitude shell encoded by color, generation
  encoded by marker shape; diagonal panels show marginal histograms.}
  \label{fig:constellation-map}
  \script{fig_constellation_map.py}
\end{figure}

The locked sweep retired with $24{,}641$
$(\mathrm{TLE}_{i},\,\mathrm{TLE}_{j})$ pairs distributed across $501$
Starlink satellites and the four target staleness offsets
$\Delta t \in \{6,\,24,\,72,\,168\}$~h; the GMAT manifest reports
$24{,}641 / 24{,}641$ propagations completed with zero integration or
preprocessing failures. The per-(altitude shell $\times$ generation)
composition of the surviving corpus is the one already given in
Table~\ref{tab:corpus}, and the geometric coverage of the sampled
satellites in the three Keplerian shape elements $(a,\,e,\,i)$ is shown
in Figure~\ref{fig:constellation-map}: the three near-vertical bands
in semi-major axis reproduce the operational altitude shells to within
the per-satellite station-keeping band; eccentricity is bounded at
$\lesssim 4\times 10^{-4}$ across the entire sample (the operational
near-circular signature of an actively phase-controlled fleet); and
inclination clusters at the documented $43^{\circ}$, $53^{\circ}$,
$70^{\circ}$, and $97^{\circ}$ deployment lanes, with v1.0 confined to
the $53^{\circ}$ Group~1 shells, v1.5 spanning $43^{\circ}$ and
$53^{\circ}$, and v2-mini populating the newer $43^{\circ}$,
$70^{\circ}$, and $97^{\circ}$ shells. The geometric narrowness of the
corpus in $(a,\,e)$ matters downstream: the per-cell comparisons
reported in Sections~\ref{sec:results-h1}--\ref{sec:results-h3}
contrast propagator skill at near-identical orbit geometry, so any
inter-cell differences are attributable to atmosphere and platform
properties rather than to gross variation in dynamical regime.

We organise the reading of Section~\ref{sec:results} around the three
hypotheses of Section~\ref{sec:introduction}. Pooled across the
$24{,}641$ corpus pairs, the L$_{2}$ position error against the
next-TLE proxy grows from a median of $0.94$~km (SGP4) /
$1.49$~km (high-fid) at the $6$~h horizon to $11.7$ / $24.5$~km at
$3$~d and $38.5$ / $76.0$~km at $7$~d. Section~\ref{sec:results-h1}
parametrises that growth into the staleness curves of H1;
Section~\ref{sec:results-h2} contrasts the two propagators pair by
pair (H2); Section~\ref{sec:results-error-structure} resolves the
position-error vector into its radial, along-track, and cross-track
components; and Section~\ref{sec:results-h3} regresses the per-sat
staleness coefficient against the daily F10.7 flux to test H3.

\subsection{Staleness curves and power-law fits (H1)}
\label{sec:results-h1}

\begin{figure}[ht]
  \centering
  \includegraphics[width=\linewidth]{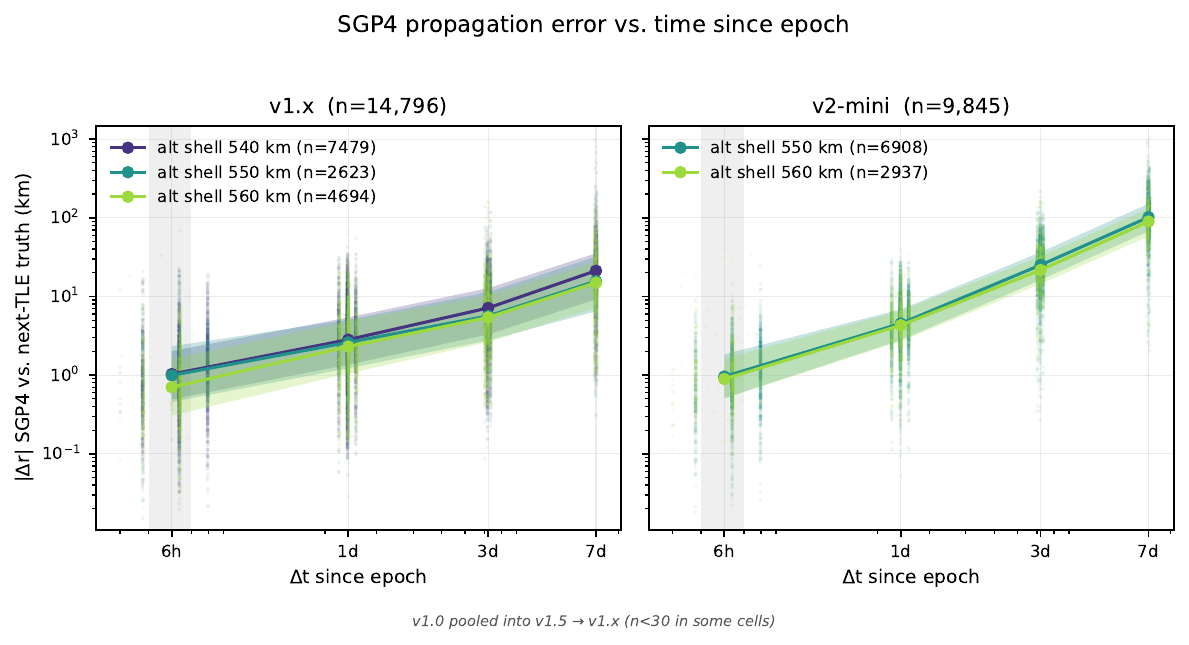}
  \caption{SGP4 propagation error against the next-TLE proxy vs.
  $\Delta t$ since epoch, by altitude shell within
  pooled-generation panels. Light scatter shows individual
  $(\Delta t, |\Delta\mathbf{r}|)$ pairs; bold markers connected by lines
  give the per-bucket median and the shaded band gives the
  $25^{\mathrm{th}}$--$75^{\mathrm{th}}$ percentile range. Log-log axes;
  identical $y$-limits to Figure~\ref{fig:hifi-growth}. The light grey
  vertical band marks the consumer-relevant 6~h horizon (the
  $\pm 2$~h pair-matching tolerance,
  Section~\ref{sec:methods-pairs}); this bucket is floor-limited
  per Section~\ref{sec:background-truth-floor}.}
  \label{fig:sgp4-growth}
  \script{fig_sgp4_growth.py}
\end{figure}

\begin{figure}[ht]
  \centering
  \includegraphics[width=\linewidth]{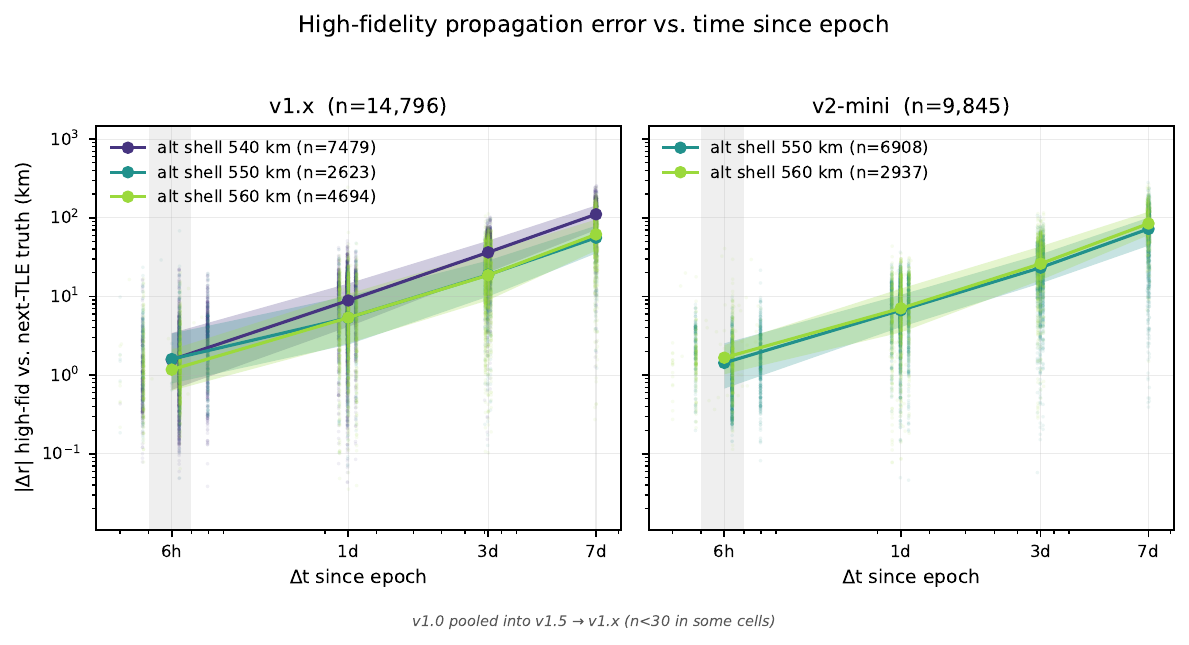}
  \caption{High-fidelity propagation error vs. $\Delta t$ since epoch,
  rendered with identical layout and axes to Figure~\ref{fig:sgp4-growth}
  for direct visual comparison. The 6~h consumer-horizon band is
  shaded as in Figure~\ref{fig:sgp4-growth}.}
  \label{fig:hifi-growth}
  \script{fig_hifi_growth.py}
\end{figure}

Figures~\ref{fig:sgp4-growth} and~\ref{fig:hifi-growth} show the
per-bucket median position error and its
$25^{\mathrm{th}}$--$75^{\mathrm{th}}$ percentile band, by altitude
shell, on identical log-log axes and identical $y$-limits so the
SGP4-vs-high-fid visual comparison is honest. Within the v1.x panel
the curves separate mildly with altitude (the $540$~km shell carries
the largest median at every $\Delta t$, consistent with drag dominance
at the lowest altitude in the corpus); within the v2-mini panel the
two populated shells collapse, the dominant offset being inter-cohort
rather than inter-shell. Walking the v1.x cells from $\Delta t = 6$~h
upwards, the SGP4 median grows from $1.01$~km
(550~km shell) to $2.60$, $5.64$, and $15.67$~km at $1$, $3$, and
$7$~d respectively; the high-fidelity arm tracks the same curve
shape at the same shell but sits a factor of $1.6$--$3.6$ higher
across the four buckets, and is uniformly above SGP4 on v1.x across
the corpus. The pattern partially carries over to v2-mini --- SGP4
itself grows more steeply ($0.96 \to 4.53 \to 25.11 \to 101.71$~km on
the SGP4 arm at the $550$~km shell) --- but the ordering of the two
medians flips with $\Delta t$: the high-fidelity arm sits above SGP4
at $6$~h and $1$~d but falls \emph{below} SGP4 at $3$~d ($23.3$
vs.\ $25.1$~km) and $7$~d ($72.3$ vs.\ $101.7$~km) on the $550$~km
v2-mini cell, and at $7$~d ($84.5$ vs.\ $89.8$~km) on the $560$~km
v2-mini cell. The cohort-level pattern from the per-bucket
aggregates --- high-fid uniformly above SGP4 on v1.x, crossing over
below SGP4 on the long-$\Delta t$ v2-mini cells --- previews the
pair-by-pair H2 reading of Section~\ref{sec:results-h2}.

\begin{figure}[ht]
  \centering
  \includegraphics[width=\linewidth]{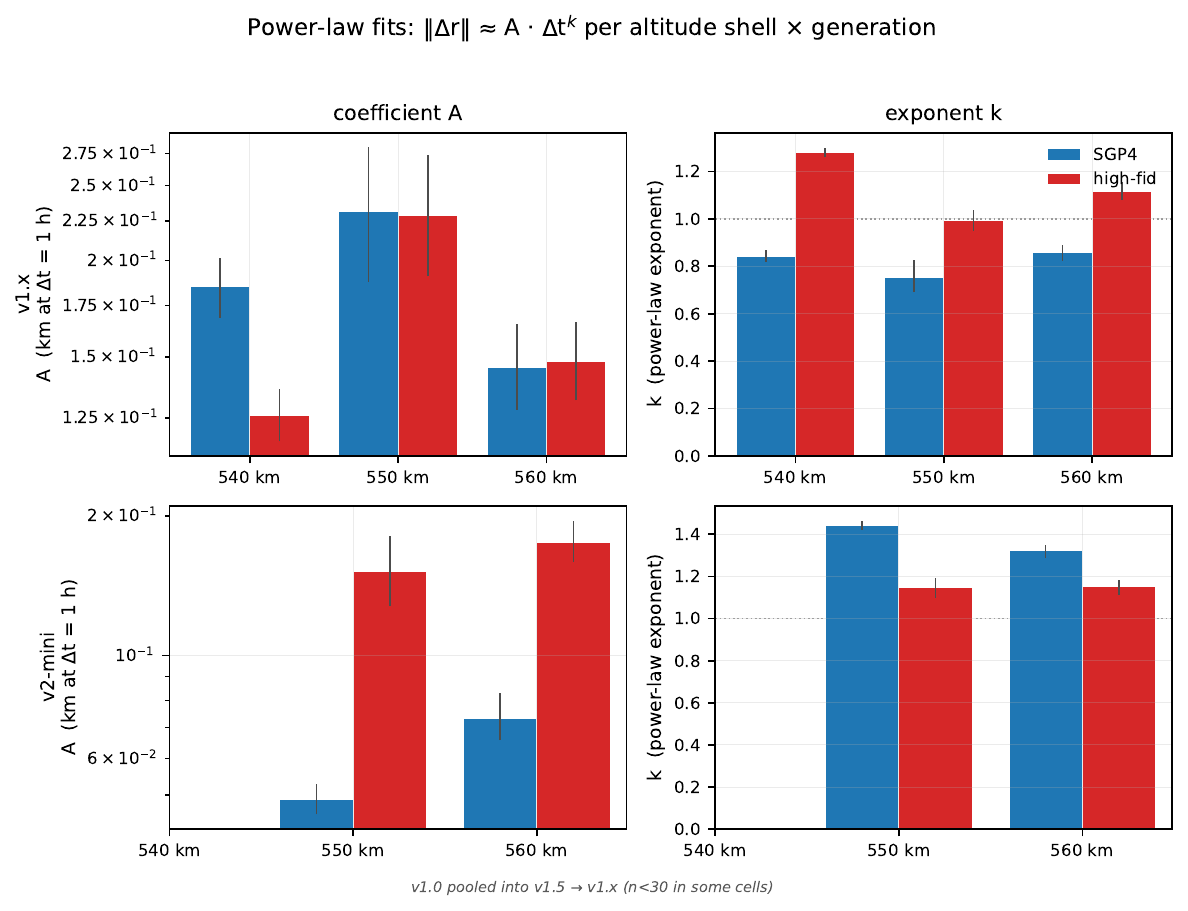}
  \caption{Power-law fits $\|\Delta\mathbf{r}\| \approx A \cdot \Delta t^{k}$
  per (altitude shell $\times$ pooled generation) cell. Bars show the
  point estimate; vertical lines give $95\%$ bootstrap percentile CIs
  from $1{,}000$ resamples drawing satellites with replacement and
  pooling all of each drawn satellite's pairs in the cell, per the
  estimator specification in Section~\ref{sec:methods-statistics}. The
  visual tightness of the CI bars on the well-populated cells reflects
  the $n \approx 167$ satellites per shell, the geometric narrowness of
  the Starlink shells in $(a, e, i)$, and log-axis compression of the
  per-pair scatter --- not under-accounting for within-satellite
  correlation, which the satellite-level resample is constructed
  specifically to absorb (Section~\ref{sec:methods-statistics}); the
  parametric mixed-effects cross-check of
  Appendix~\ref{app:mixed-effects} returns slopes inside the
  bootstrap CIs on every populated cell. Left subpanel per row: coefficient
  $A$ on a log axis; right subpanel: exponent $k$ with $k = 1$ marked.
  Numerical values are listed in Table~\ref{tab:powerlaw}.}
  \label{fig:powerlaw-fits}
  \script{fig_powerlaw_fits.py}
\end{figure}

The curves are well-fit by the H1 form
$\lVert\Delta\mathbf{r}(\Delta t)\rVert \approx A\,\Delta t^{k}$ at
the per-pair level (Section~\ref{sec:methods-statistics}).
Figure~\ref{fig:powerlaw-fits} renders the fitted parameters per
(altitude shell $\times$ pooled generation) cell with $95\%$
satellite-level bootstrap CIs; the numerical values are listed in
Table~\ref{tab:powerlaw}. Three patterns survive the bootstrap.
First, on the v1.x cohort SGP4 yields a sub-linear exponent
($\hat k$ ranges from $0.75$~$[0.69,\,0.83]$ at $550$~km through
$0.84$~$[0.82,\,0.87]$ at $540$~km to $0.86$~$[0.82,\,0.89]$ at
$560$~km), while the high-fidelity arm on the same cohort spans
$0.99$~$[0.95,\,1.04]$ at $550$~km (where $\hat k$ sits at the
$k=1$ boundary and the CI straddles it), $1.12$~$[1.08,\,1.15]$
at $560$~km, and $1.28$~$[1.26,\,1.30]$ at $540$~km. The likelihood-ratio test
rejects both the $k = 1$ and $k = 2$ nulls at $p < 10^{-3}$ in every
populated cell except the high-fidelity v1.x cell at $550$~km, where
the $k=1$ null is not rejected ($p_{k=1} = 0.76$,
Table~\ref{tab:powerlaw}) and the cell sits in the mean-motion-bias
limit of Section~\ref{sec:methods-powerlaw-physics}. Elsewhere the
mixture interpretation, with $\hat k$ encoding the relative weight
of a $B^{\!\star}$ mean-motion bias versus a constant unmodelled
along-track acceleration, is the physically consistent reading.
Second, the two propagators trade $A$ for $k$ in opposite directions
across the v1.x cells: high-fid carries the smaller (or equal)
coefficient at the $\Delta t = 1$~h normalisation point (the
bolded entries of Table~\ref{tab:powerlaw}) but the steeper slope,
so the two staleness curves cross within the corpus window and the
v1.x gap of Figures~\ref{fig:sgp4-growth}--\ref{fig:hifi-growth}
widens with $\Delta t$. Third, the v2-mini cohort is super-linear
under \emph{both} propagators ($\hat k$ between $1.15$ and $1.44$),
with the unusual feature that SGP4 carries the smaller coefficient
\emph{and} the steeper slope at both shells, so on v2-mini it is the
high-fidelity arm that becomes the slower-growing curve and the two
staleness curves cross with SGP4 climbing past high-fid by $3$~d ---
the §\ref{sec:results-h1}-paragraph statement of the v2-mini
crossover that drives the H2 reading in
Section~\ref{sec:results-h2}.

\begin{table}[ht]
  \centering
  \caption{Power-law fit parameters per (altitude shell $\times$ pooled
  generation), for SGP4 and high-fidelity propagators. Values for the
  coefficient $A$ and exponent $k$ are point estimates from the
  per-pair OLS fit (Section~\ref{sec:methods-statistics}) followed by
  the $95\%$ satellite-level bootstrap CI. $R^{2}$ is the
  coefficient of determination of the unconstrained per-pair fit;
  the $p_{k=1} / p_{k=2}$ column reports the likelihood-ratio-test
  p-values for the two slope nulls $k = 1$ and $k = 2$. Per row, the
  $A$ entry of whichever propagator has the smaller fitted coefficient
  is bolded; this is the propagator with the lower error at the
  $\Delta t = 1\,\text{h}$ normalisation point, not necessarily at
  the longer bucket centres.}
  \label{tab:powerlaw}
  \resizebox{\textwidth}{!}{%
  \input{tables/tab_powerlaw.tex}\unskip\label{tables/tab_powerlaw.tex}\unskip%
}
  \script{fig_powerlaw_fits.py}
\end{table}

H1 is therefore not merely supported but \emph{quantified}: every
populated cell of the corpus carries a statistically distinguishable
power-law staleness curve, the per-cell $(A,\,k)$ pair varies
systematically and interpretably across (shell $\times$ generation),
and the cohort-specific values of Table~\ref{tab:powerlaw} are the
practitioner-facing summary the analysis is built to provide.
The behaviour above the truth-floor regime of
Section~\ref{sec:background-truth-floor} -- $\Delta t \gtrsim 1$~d, where
the observed error climbs into double-digit kilometres and the
OD-residual contribution becomes percent-level -- carries the
load-bearing fits; the $6$~h cell is reported alongside for the
operational completeness already documented in §\ref{sec:methods-pairs}
but is interpreted in
Section~\ref{sec:discussion} against the
Section~\ref{sec:background-truth-floor} floor.

\subsection{Pair-by-pair propagator comparison (H2)}
\label{sec:results-h2}

\begin{figure}[ht]
  \centering
  \includegraphics[width=\linewidth]{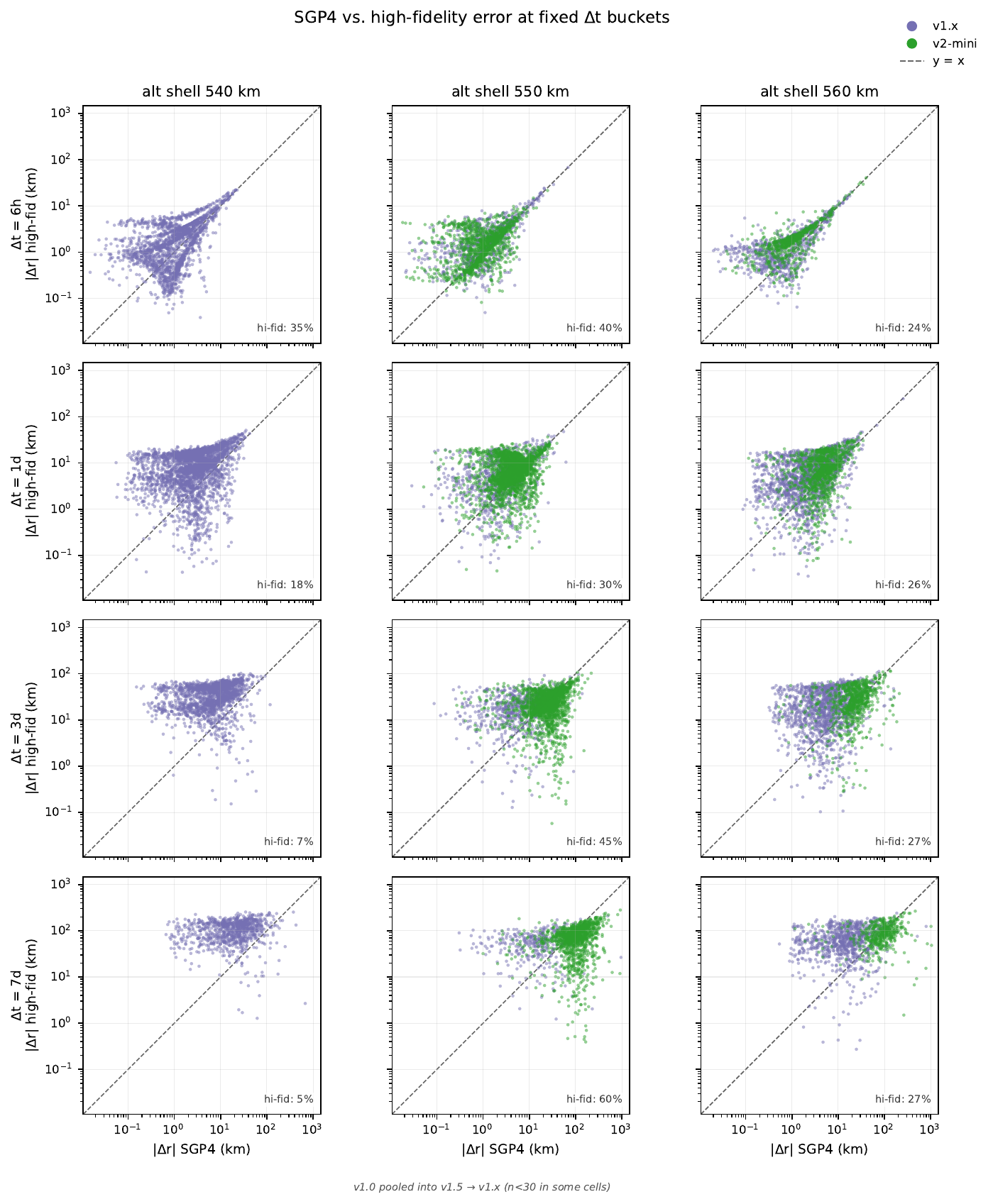}
  \caption{SGP4 vs.\ high-fidelity error per pair, at fixed $\Delta t$
  buckets (rows) and altitude shells (columns). Color encodes pooled
  generation; the dashed $y = x$ line marks parity, so points below are
  pairs where the high-fidelity model beat SGP4 in 3D L$_{2}$ norm. The
  per-panel label in each lower-right corner reports the fraction of
  cell pairs below the diagonal (``hi-fid wins''); a bootstrap-quantified
  per-cell version of the same fraction (plus an along-track-only
  variant) is in Table~\ref{tab:propagator-wins}. A hexbin density
  variant of the same scatter is in
  Appendix~\ref{app:propagator-scatter-hexbin}
  (Figure~\ref{fig:propagator-scatter-hexbin}).}
  \label{fig:propagator-scatter}
  \script{fig_propagator_scatter.py}
\end{figure}

The aggregate statistics of Section~\ref{sec:results-h1} compare
medians; H2 asks the sharper pair-wise question: in what fraction of
$(\mathrm{TLE}_{i},\,\mathrm{TLE}_{j})$ pairs is the high-fidelity
$|\Delta\mathbf{r}|$ smaller than the SGP4 $|\Delta\mathbf{r}|$ against
the same proxy truth? Figure~\ref{fig:propagator-scatter} plots every
pair on the $(|\Delta\mathbf{r}_{\mathrm{SGP4}}|,\,
|\Delta\mathbf{r}_{\mathrm{hifi}}|)$ plane, broken out by altitude
shell (columns) and $\Delta t$ bucket (rows). Each panel's lower-right
annotation reports the fraction of pairs sitting below the $y = x$
diagonal (``hi-fid wins''); the bootstrap-quantified per-cell version
of that fraction, together with the along-track-only variant
$|\Delta r_{\mathrm{along},\,\mathrm{hifi}}| <
|\Delta r_{\mathrm{along},\,\mathrm{SGP4}}|$ that
Figure~\ref{fig:error-decomposition} below motivates as the natural
metric in the long-$\Delta t$ regime, is given in
Table~\ref{tab:propagator-wins}.

\begin{table}[ht]
  \centering
  \caption{Per-cell fraction of $(\mathrm{TLE}_{i},\,\mathrm{TLE}_{j})$
  pairs on which the high-fidelity propagator achieves a smaller
  position error than SGP4 against the next-TLE proxy. The 3D
  L$_{2}$ column reads ``$\lVert\Delta\mathbf{r}_{\mathrm{hifi}}\rVert
  < \lVert\Delta\mathbf{r}_{\mathrm{SGP4}}\rVert$'' (the metric
  underlying Figure~\ref{fig:propagator-scatter}); the along-track
  column reads
  ``$\lvert\Delta r_{\mathrm{along},\,\mathrm{hifi}}\rvert <
  \lvert\Delta r_{\mathrm{along},\,\mathrm{SGP4}}\rvert$'' on the
  RSW component of Figure~\ref{fig:error-decomposition}. Both
  fractions carry $95\%$ satellite-level percentile bootstrap CIs
  ($1{,}000$ resamples drawing satellites with replacement and pooling
  all of each drawn satellite's pairs in the cell; identical
  resampling structure to Table~\ref{tab:powerlaw}). Per-cell rows
  with fewer than $30$ pairs are suppressed; pooled-per-$\Delta t$
  rows aggregate across all shells and generations. The
  cohort-resolved decomposition of these rows (per-$(\text{shell}
  \times \Delta t \times \text{generation})$ win fractions) is in
  Table~\ref{tab:propagator-wins-by-gen} alongside the
  Section~\ref{sec:discussion-v2mini} cohort-mechanism discussion.}
  \label{tab:propagator-wins}
  \input{tables/tab_propagator_wins.tex}
\end{table}

Pooled across all cells, the high-fidelity arm beats SGP4 on
$33.7\%$~$[32.2,\,35.3]$ of $6$~h pairs in 3D L$_{2}$ norm,
$25.1\%$~$[23.9,\,26.1]$ at $1$~d, $28.7\%$~$[26.6,\,30.9]$ at
$3$~d, and $35.3\%$~$[31.9,\,38.8]$ at $7$~d. The non-monotone
pooled trajectory --- worst at $1$~d, recovering toward the $6$~h
fraction by $7$~d --- masks a strong per-cell pattern that
Table~\ref{tab:propagator-wins} resolves below. The along-track
variant sits within sampling error of the L$_{2}$ variant at every
horizon except $6$~h, where it climbs to $36.0\%$~$[34.3,\,37.6]$:
at the shortest horizon the cross-track and radial components of
$\Delta\mathbf{r}$ are still appreciable contributions to the
L$_{2}$ norm (see Section~\ref{sec:results-error-structure} below),
while at $\geq 1$~d the L$_{2}$ comparison is already an
along-track comparison in disguise, and the two metrics return the
same fraction within the bootstrap CI. Either way SGP4 wins on
$\sim\!65$--$75\%$ of pairs across the corpus, which is the
per-pair statement of the H2 negative finding.

The per-cell decomposition exposes a strong cohort dependence behind
the pooled numbers. At $540$~km the high-fidelity-wins fraction
collapses from $34.6\%$~$[32.3,\,36.8]$ at $6$~h to
$5.4\%$~$[3.9,\,7.0]$ at $7$~d: as $\Delta t$ grows the high-fidelity
arm at the lowest shell saturates against a propagation floor
visible in Figure~\ref{fig:propagator-scatter} as the
``$|\Delta\mathbf{r}_{\mathrm{hifi}}|$ horizontal band'' centred near
$100$--$200$~km in the $3$~d and $7$~d panels, while the SGP4
distribution remains spread over three decades. That signature is
the joint fingerprint of a near-constant force-model bias on the
high-fidelity arm (drag mismodelling on the lowest-altitude cohort)
and a wider operator-OD-residual distribution on SGP4 -- the
combination yields a stripe of paired points whose ordinate is
nearly $\Delta t$-determined and whose abscissa varies. At $550$~km
the shell-level high-fidelity-wins fraction is qualitatively different:
it climbs with $\Delta t$ from $39.8\%$~$[37.0,\,42.6]$ at $6$~h to
$60.2\%$~$[55.9,\,64.4]$ at $7$~d, the only shell-level cell of the
corpus where the high-fidelity arm beats SGP4 on a majority of pairs.
At $560$~km the comparable shell-level wedge is weaker --- the cell's
$7$~d high-fidelity-wins fraction is $26.7\%$~$[22.1,\,31.6]$ ---
suggesting at first reading that the $550$~km flip is an isolated
exception.

That reading is partly an artefact of cohort blending in the
shell-level statistic. The $550$~km $\times$ $7$~d cell is
$78\%$ v2-mini by pair count whereas the $560$~km $\times$ $7$~d
cell is only $34\%$ v2-mini, so the same per-cohort behaviour
projects very differently onto the shell-level wins fraction when
the v1.5 majority drags the $560$~km number down. The
cohort-resolved fractions in
Table~\ref{tab:propagator-wins-by-gen} (in
Section~\ref{sec:discussion-v2mini}) make the point directly: on
the v2-mini cohort alone the high-fidelity arm wins
$72.6\%$~$[69.9,\,75.4]$ of $7$~d pairs at $550$~km \emph{and}
$55.1\%$~$[49.9,\,60.4]$ of $7$~d pairs at $560$~km, both clearly
above the $50\%$ null; while on v1.x the arm loses badly at
both shells ($16.1\%$~$[11.0,\,22.2]$ at $550$~km;
$11.9\%$~$[8.8,\,15.5]$ at $560$~km). The cohort signal at
$7$~d is therefore cohort-uniform, not shell-specific. The same
table shows the v2-mini hi-fid-wins fraction rising with
$\Delta t$ at \emph{both} populated shells from below half at
$6$~h to a majority at $7$~d.
Section~\ref{sec:discussion-v2mini} returns to the physical
mechanism behind the cohort split.

H2 therefore holds with the reading: the high-fidelity propagator
initialised from public TLEs does not improve on SGP4 in the bulk
of the corpus, and the pooled win fraction is statistically distant
from the $50\%$ null at every $\Delta t$. The cohort-specific
exception --- v2-mini at long $\Delta t$, present at both $550$ and
$560$~km --- is the regime where the SGP4 cohort itself struggles
most (super-linear $\hat k = 1.44$ and $1.32$ respectively in
Table~\ref{tab:powerlaw}), making even an inherently biased
high-fidelity propagator competitive on a non-trivial fraction of
pairs. Section~\ref{sec:discussion} decomposes the bulk negative
result into three distinct mechanisms, and
Section~\ref{sec:discussion-v2mini} elaborates the cohort-resolved
reading.

\subsection{Error decomposition}
\label{sec:results-error-structure}

\begin{figure}[ht]
  \centering
  \includegraphics[width=\linewidth]{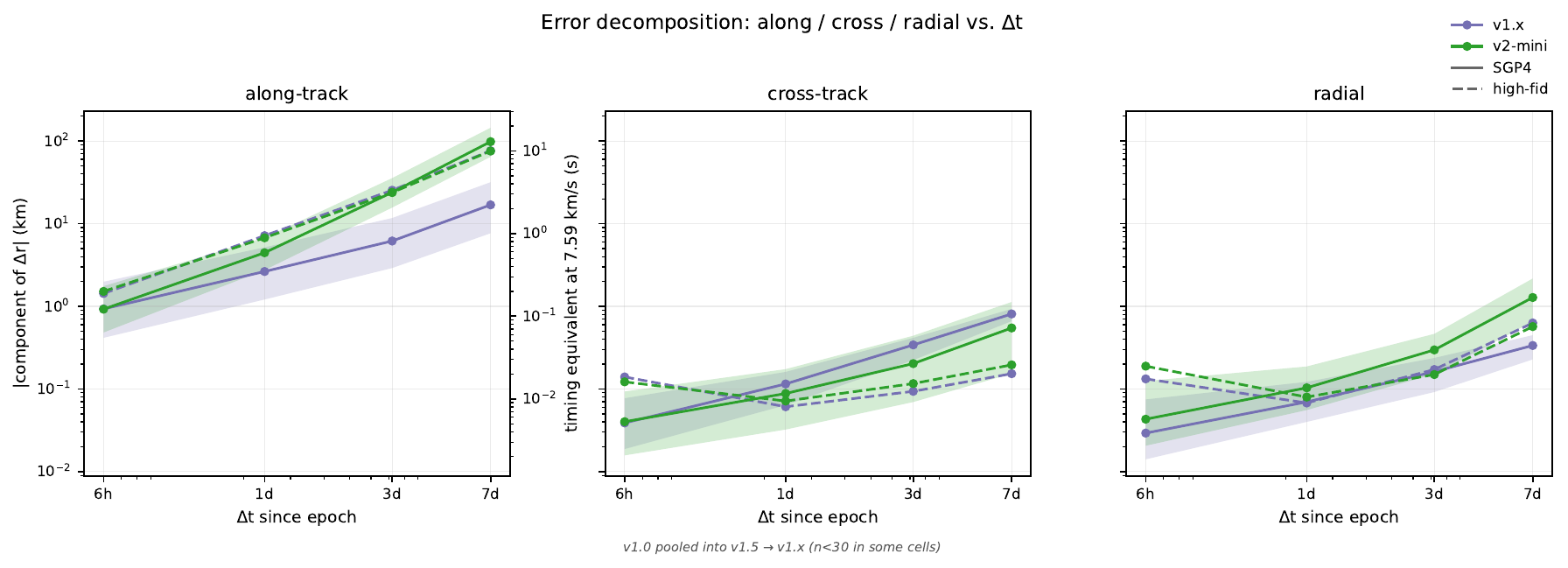}
  \caption{Error decomposition into along-track, cross-track, and radial
  components of $\Delta\mathbf{r}$ in the next-TLE-proxy RSW frame.
  Per-bucket medians with IQR ribbons; solid lines are SGP4, dashed
  lines are high-fidelity; color encodes pooled generation. The
  along-track panel carries a secondary $y$-axis in seconds of
  orbital phase, converted by $\Delta t_{\rm{phase}} = \Delta s / v_{c}$
  at the $550$~km circular-orbit speed
  $v_{c} = \sqrt{\mu / (R_{\oplus} + 550\,\text{km})} \approx
  7.59$~km\,s$^{-1}$, the central-shell value.}
  \label{fig:error-decomposition}
  \script{fig_error_decomposition.py}
\end{figure}

Figure~\ref{fig:error-decomposition} resolves $\Delta\mathbf{r}$ into
the standard radial, along-track, and cross-track components in the
next-TLE-proxy RSW frame (Section~\ref{sec:methods-metrics}). The
along-track median sits a clean two decades above the cross-track
and radial medians at every $\Delta t$ bucket and in every populated
cohort, and the gap widens with $\Delta t$. On the v1.x cohort,
along-track grows roughly $1 \to 4 \to 6 \to 16$~km from $6$~h to
$7$~d on the SGP4 arm; the v2-mini cohort grows $1 \to 4 \to 24 \to
98$~km on the same arm and reaches $\sim 76$~km at $7$~d on the
high-fidelity arm. Over the same horizons the cross-track and
radial medians grow from $\mathcal{O}(0.03$--$0.1)$~km at $6$~h to
$\mathcal{O}(0.3$--$1)$~km at $7$~d, two decades below the
along-track band throughout. By $\Delta t = 7$~d, the L$_{2}$ norm
of Figures~\ref{fig:sgp4-growth}--\ref{fig:hifi-growth} and the
along-track component of Figure~\ref{fig:error-decomposition} are
indistinguishable within the line widths of the figure: at the
long-staleness limit, the L$_{2}$ position error \emph{is} the
along-track error.

The decomposition gives the
along-track-norm in operationally meaningful units:
$1$~km of along-track displacement at the $550$~km shell's circular
speed $v_{c} \approx 7.59$~km\,s$^{-1}$ corresponds to
$\sim 0.13$~s of orbital phase, marked on the secondary axis of the
along-track panel. The $6$~h headline median of $\sim 1$~km from
Section~\ref{sec:results-h1} is therefore a sub-orbit-second timing
miss along the velocity direction, while the $7$~d high-fidelity
median of $\sim 76$~km corresponds to $\sim 10$~s of phase.
Cross-track and radial components -- bounded under the dominant LEO
perturbations by orbital-period oscillation and second-order coupling
\citep[\S\,3.4]{vallado2013} -- accumulate freely over neither
horizon, and the SGP4 and high-fidelity dashed lines on those two
panels overlay each other within the IQR ribbons. The H2 propagator
gap is therefore an along-track gap; the discussion in
Section~\ref{sec:discussion-along-track} returns to why.

\subsection{Solar modulation (H3)}
\label{sec:results-h3}

\begin{figure}[ht]
  \centering
  \includegraphics[width=\linewidth]{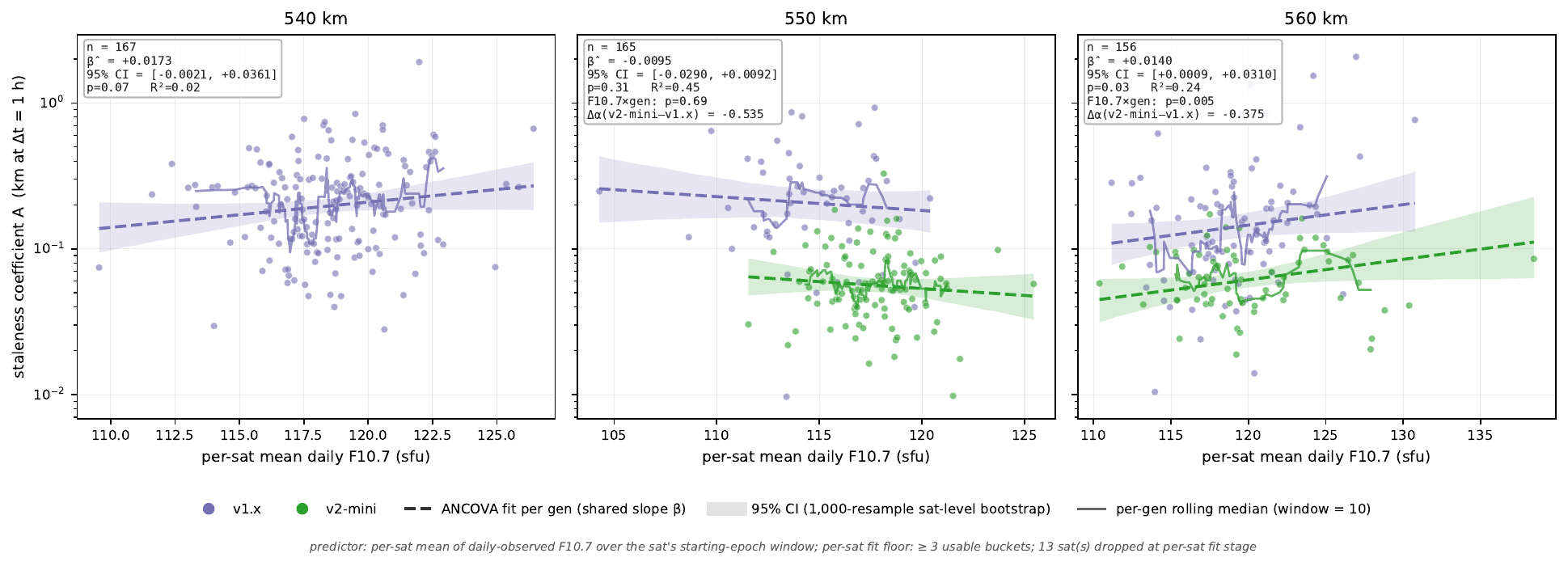}
  \caption{Per-satellite SGP4 staleness coefficient $A$ vs.\ the
  mean daily-observed F10.7 flux over the satellite's corpus time
  window, one panel per altitude shell ($540$, $550$, $560$~km).
  Markers are coloured by pooled generation (purple v1.x, green
  v2-mini); the $540$~km shell has no v2-mini satellites in the
  corpus (Table~\ref{tab:corpus}). Within each panel: per-satellite
  scatter, a per-generation rolling median (solid) for visual
  context, and the ANCOVA fit of Eq.~\eqref{eq:h3-model} drawn as
  one dashed OLS line per generation present, with $95\%$ percentile
  CI ribbons. The two lines per panel share the fitted slope
  $\hat\beta$ and differ only in the per-generation intercept
  $\hat\alpha_{\mathrm{gen}}$; ribbons combine the slope and
  intercept-offset bootstrap distributions consistently. The
  in-panel annotation reports the satellite count, $\hat\beta$, its
  $95\%$ bootstrap CI, the $t$-stat $p$-value for
  $H_{0}\!: \beta = 0$, the model $R^{2}$, the partial-$F$
  $p$-value for the interaction-vs-additive comparison
  (where the shell carries more than one cohort), and the
  intercept offset
  $\hat\alpha_{\mathrm{v2\text{-}mini}} -
  \hat\alpha_{\mathrm{v1.x}}$.}
  \label{fig:solar-modulation}
  \script{fig_solar_modulation.py}
\end{figure}

H3 asks whether the per-satellite SGP4 staleness coefficient $A_{i}$
of Section~\ref{sec:methods-statistics} tracks the daily-observed
F10.7 flux in the most drag-sensitive cohorts.
Figure~\ref{fig:solar-modulation} plots the per-sat $A_{i}$ scatter
against the per-sat mean of the CelesTrak daily F10.7 over the
satellite's window of starting epochs, one panel per altitude shell,
and overlays the additive ANCOVA fit of Eq.~\eqref{eq:h3-model}
(one dashed line per generation present, sharing the fitted slope
$\hat\beta$ across cohorts and absorbing inter-cohort offsets into
$\hat\alpha_{\mathrm{gen}}$ as predicted by
Eq.~\eqref{eq:h3-slope-altitude-only}).

The headline daily-F10.7 ANCOVA fit returns one slope per shell, summarised in
Table~\ref{tab:h3-fits}: a positive slope at $540$~km that does not
exclude zero at the corpus's narrow predictor span ($\sim 17$~sfu
across satellites); a slope indistinguishable from zero at $550$~km;
and a positive slope at $560$~km that clears conventional
significance. The $560$~km cell is therefore the only one where H3
modulation is statistically distinct from the null on the present
30-day window. The $540$~km cell is direction-consistent with the
textbook positive density gradient with F10.7 at LEO, but the
intra-corpus F10.7 lever is too narrow to discriminate the
predicted slope from zero on its own.

\begin{table}[ht]
  \centering
  \caption{Per-shell ANCOVA fit of
  Eq.~\eqref{eq:h3-model} with the daily-observed F10.7 as
  predictor: slope estimate $\hat\beta$ in $\text{sfu}^{-1}$, $95\%$
  satellite-level percentile bootstrap CI, two-sided $t$-stat
  $p$-value for $H_{0}\!:\beta = 0$, and model $R^{2}$. The
  partial-$F$ interaction test column reports the $p$-value of
  the F10.7$\times$generation interaction against the additive
  restriction (shells with a single cohort have no such test).
  The intercept-offset column reports
  $\hat\alpha_{\mathrm{v2\text{-}mini}} - \hat\alpha_{\mathrm{v1.x}}$
  for shells where both cohorts are present.}
  \label{tab:h3-fits}
  \begin{tabular}{lrcrrrr}
    \toprule
    Shell & $n_{\mathrm{sats}}$ & cohorts & $\hat\beta$ & $95\%$ CI &
    $p$ & $\Delta\hat\alpha$ \\
    (km) & & & (sfu$^{-1}$) & (sfu$^{-1}$) & & ($\log_{10}$) \\
    \midrule
    540 & 167 & v1.x & $+0.017$ &
      $[-0.002,\,+0.036]$ & $0.07$ & --- \\
    550 & 165 & v1.x, v2-mini & $-0.009$ &
      $[-0.029,\,+0.009]$ & $0.31$ & $-0.54$ \\
    560 & 156 & v1.x, v2-mini & $+0.014$ &
      $[+0.001,\,+0.031]$ & $0.034$ & $-0.38$ \\
    \midrule
    \multicolumn{3}{l}{Interaction-vs-additive partial-$F$}
      & \multicolumn{2}{l}{F10.7$\times$gen} & $p$-value & \\
    \midrule
    540 & & & & & --- & \\
    550 & & & \multicolumn{2}{l}{additive ok} & $0.69$ & \\
    560 & & & \multicolumn{2}{l}{additive rejected} & $0.005$ & \\
    \bottomrule
  \end{tabular}
\end{table}

The $560$~km interaction test does reject the additive restriction
($p = 0.005$), indicating a cohort-specific F10.7 response at that
shell that Eq.~\eqref{eq:h3-slope-altitude-only} does not predict.
The departure is flagged in Section~\ref{sec:discussion} as either
a real second-order effect of the v2-mini bus or a window-end
sampling pattern interacting with the F10.7 trend over the analysis
window; the additive headline slope at $560$~km remains the
$\hat\beta$ that
Eq.~\eqref{eq:h3-slope-altitude-only} concerns directly, and we
report it without substitution by the interaction model.

The $81$-day-centred robustness fits, reported per-shell in
Section~\ref{sec:methods-h3-regression}, return slopes that are
physically wrong-signed at $550$ and $560$~km (negative on a quantity
whose thermospheric-density gradient is uniformly positive at LEO)
and reach nominal significance only because the predictor's
$\sim 1.3$~sfu span across satellites is dominated by where in
April~2026 each satellite's starting epochs cluster rather than by a
genuine solar-driver signal. We treat those fits as the
limit-of-leverage diagnostic they are -- documented in
Section~\ref{sec:methods-h3-regression} alongside the
daily-mean fit -- and the daily-mean $\hat\beta$ of
Figure~\ref{fig:solar-modulation} as the load-bearing H3 report on
the present 30-day corpus. H3 is therefore detectable in direction
across the two cells where the analysis has discriminating power
(positive $\hat\beta$ at $540$ and $560$~km, the two shells with the
broadest within-cohort F10.7 lever) but is statistically distinct
from zero only at one of them.

\section{Discussion}
\label{sec:discussion}

The empirical headline of Section~\ref{sec:results} --- that the
high-fidelity propagator initialised from a public TLE is behind
SGP4 against the next-TLE proxy at every shell-pooled $\Delta t$
bucket of the corpus, losing on $65$--$75\%$ of pairs depending on
horizon (Table~\ref{tab:propagator-wins}) --- supports a more
specific reading than ``high-fidelity propagation is wasted''. The
load-bearing statement is the conditional one: \emph{higher
physical fidelity in the propagator does not imply superior
predictive accuracy against next-TLE-derived truth when the initial
condition is itself a state estimated under a different dynamical
model}. The H2 gap is then not a property of the high-fidelity
force model in isolation but of the joint
(initial-state, truth-construction, force-model) triple that the
public-TLE workflow imposes. Section~\ref{sec:discussion-h2-mechanisms}
walks that joint object through three mechanisms whose dominance
regimes partition the four $\Delta t$ buckets cleanly between them;
Section~\ref{sec:discussion-v2mini} resolves the bulk negative
result by cohort and isolates the v2-mini long-$\Delta t$ regime
where high-fidelity propagation overtakes SGP4 on a majority of
pairs; the along-track structure of the error budget
(Section~\ref{sec:discussion-along-track}), the practitioner-facing
quantitative verdict (Section~\ref{sec:discussion-practical}), and
the limitations and follow-ups
(Sections~\ref{sec:discussion-covariance}--\ref{sec:discussion-extensions})
follow.

\subsection{Three mechanisms behind the H2 gap}
\label{sec:discussion-h2-mechanisms}

\textbf{Mechanism 1 --- Operator-OD residual dominance at short
$\Delta t$.} A useful sanity check on the $6$~h headline number is to
compare it directly with the truth-floor diagnostic of
Section~\ref{sec:background-truth-floor}. Pooled across the eight
cohort representatives the diagnostic returns a median L$_{2}$
residual of $1.11$~km on a $\sim 5$~h short arc (per-cohort medians
$0.81$--$1.37$~km, Table~\ref{tab:truth-floor}), whereas the SGP4
pooled median at $\Delta t = 6$~h is below that value at $\sim 0.94$~km
(Section~\ref{sec:results-h1}). Two TLE-derived states separated by
one operator update interval cannot agree better than the floor
implied by their independent OD residuals; the only way the observed
$6$~h SGP4 number sits beneath the diagnostic floor is by partial
cancellation between adjacent operator OD residuals --- shared attitude
assumptions, shared recent-maneuver history, and common-mode
atmospheric mis-fits within the operator's batch window all act as
correlated contributions to $\Delta\mathbf{r}_{\mathrm{OD}}(t_{i})$
and $\Delta\mathbf{r}_{\mathrm{OD}}(t_{j})$ at consecutive epochs.
The gap between $1.11$~km and $0.94$~km is then a coarse measurement
of how much of the $6$~h signal is floor versus propagator. Treating
the diagnostic value as an estimate of the at-epoch OD residual
scale $\delta_{\mathrm{OD}}$ and propagating it through the
power-law fit $\lVert\Delta\mathbf{r}\rVert \approx A\,\Delta t^{k}$
of Table~\ref{tab:powerlaw}, the $\delta_{\mathrm{OD}}$ contribution
biases the fitted coefficient $A$ upwards by an amount that scales
as $\delta_{\mathrm{OD}}\,\Delta t^{-k}$ --- meaningful at $6$~h,
where it accounts for the bulk of the observed value, and negligible
by $7$~d, where the absolute observed error is two orders of
magnitude larger. Mechanism~1 therefore inflates both propagator
arms' absolute L$_{2}$ norms at the $6$~h horizon and recedes
monotonically with $\Delta t$, vanishing into the percent level at
$7$~d.

A naive expectation that floor dominance would also collapse the
$6$~h paired distribution onto the $y = x$ diagonal of
Figure~\ref{fig:propagator-scatter} is, however, wrong. Both
propagator arms inherit the \emph{same} $\Delta\mathbf{r}_{\mathrm{OD}}(t_{j})$
at the truth epoch and the same
$\boldsymbol{\Phi}(t_{j},\,t_{i})\,\Delta\mathbf{r}_{\mathrm{OD}}(t_{i})$
at the start epoch, so the OD-residual terms cancel from the
$\lVert\Delta\mathbf{r}_{\mathrm{SGP4}}\rVert -
\lVert\Delta\mathbf{r}_{\mathrm{hifi}}\rVert$ \emph{differential}
even when they dominate each absolute norm. The floor inflates the
$x$- and $y$-axes of Figure~\ref{fig:propagator-scatter} together
but does not bias the win count, and the empirical
$33.7\%$~$[32.2,\,35.3]$ hi-fid-wins fraction at $6$~h
(Table~\ref{tab:propagator-wins}) is the genuine
SGP4-vs-high-fidelity skill differential on the floor-saturated
horizon. The question that mechanism~1 \emph{does not} answer is
why that differential favours SGP4 at all; mechanism~2 supplies
the sign.

\textbf{Mechanism 2 --- Truth-construction kernel alignment.} The
next-TLE-as-proxy comparison is asymmetric in a structural way:
both the SGP4 prediction at $t_{j}$ and the truth state at $t_{j}$
are evaluated through the same analytic kernel
(SGP4 at $\mathrm{TLE}_{j}$, $\Delta t = 0$), while the
high-fidelity prediction at $t_{j}$ is evaluated through a
numerically integrated force model that lives in a different
dynamical reference (Section~\ref{sec:methods-dynamical-consistency}).
A class of systematic biases that SGP4 carries against the
underlying physical orbit --- the Brouwer mean-element formulation,
the WGS-72 gravity constants, the analytic absorption of drag into
a single fitted $B^{\!\star}$ at epoch --- is therefore mirrored
between prediction and truth on the SGP4 arm and \emph{cancels}
from $\Delta\mathbf{r}_{\mathrm{SGP4}}$. The high-fidelity arm has
no such mirror: its initial state lives on SGP4's
pseudo-osculating intermediate manifold but its forward propagation
lives on EGM2008 + NRLMSISE-00 osculating dynamics, so a
$\mathcal{O}(1)$~km Brouwer short-period mismatch
(Section~\ref{sec:methods-dynamical-consistency}) is integrated
into $\Delta\mathbf{r}_{\mathrm{hifi}}$ without compensation at the
truth construction. The scale of this kernel-alignment advantage
is exactly the $\sim 1$~km separation between the two arms at the
$6$~h horizon, and the partial decay of the advantage with
$\Delta t$ is consistent with the short-period kernel coupling
back into a secular component through resonant interaction with
mismodelled drag --- a $B^{\!\star}$ mis-fit on the SGP4 side
counteracts some of the kernel advantage at long horizons but
never enough to invert the sign of the differential.

Mechanism~2 is the explicit cost of the public-TLE workflow: a
researcher with operator-internal Cartesian state vectors would
bypass it entirely (no mean-element-to-osculating conversion, no
TEME-to-MJ2000Eq rotation, no analytic kernel imposed on the
initial condition). For consumers of public TLEs the mechanism is
unavoidable; for the present comparison it is the largest single
contribution to the H2 sign at the operationally relevant $6$~h
and $1$~d horizons.

\textbf{Mechanism 3 --- Spacecraft-property uncertainty
preferentially harms the hi-fid arm.} The third mechanism
dominates at the long-$\Delta t$ end of the corpus, where
mechanism~1 has receded to percent-level and the integrated drag
modelling error overwhelms the mechanism-2 kernel offset. At
$\Delta t = 7$~d the pooled high-fidelity L$_{2}$ median sits at
$\sim 76$~km against the SGP4 median of $\sim 38$~km
(Section~\ref{sec:results-h1}), and the corresponding pooled
hi-fid-wins fraction is $35.3\%$~$[31.9,\,38.8]$
(Table~\ref{tab:propagator-wins}) --- still below the $50\%$ null
in the bulk of the corpus, even after the long-$\Delta t$ recovery
visible in the pooled trajectory. The asymmetry is mechanical: a
systematic $C_{D} \cdot A$ bias feeds directly into the
high-fidelity drag acceleration through the $\rho\,C_{D}\,A$
product, whereas SGP4 absorbs the equivalent bias into the
operator-fitted $B^{\!\star}$ at epoch
(Section~\ref{sec:background}); the SGP4 arm is therefore
calibrated against the same data the user has access to, while
the high-fidelity arm is exposed to whatever residual error the
modeller's prior on $C_{D} \cdot A$ leaves on the table.
Appendix~\ref{app:cda-sensitivity} quantifies the effect with a
$\pm 20\%$ $C_{D} \cdot A$ perturbation on the v2-mini cohort: the
pooled median $\lvert\Delta\mathbf{r}\rvert_{\mathrm{hifi}}$
shifts by $-8.9\%$ at $0.8\times$ and $+11.0\%$ at $1.2\times$
relative to a baseline of $19.5$~km, the per-cell shift is
direction-consistent (more drag area $\Rightarrow$ larger error,
the expected sign) and peaks at $\pm 17\%$ in the
$550$~km~$\times$~$7$~d cell. At the favourable $0.8\times$ factor
--- roughly what an operator with internal mass-properties
knowledge would gain over a public-data modeller --- the
high-fidelity arm still sits $\sim 16$--$18\%$ above SGP4 in the
two most-affected v2-mini cells, so a quantitatively realistic
improvement in the modeller's $C_{D} \cdot A$ prior does not
invert the H2 sign. Spacecraft-property knowledge is therefore a
mechanism-3 lever for which the present comparison sets an
empirical upper bound on the achievable gain.

Read together, the three mechanisms partition the H2 gap by
$\Delta t$ regime: mechanism~1 dominates the absolute L$_{2}$
norms at the $6$~h horizon but cancels from the differential;
mechanism~2 carries the $6$~h--$1$~d horizons through the
kernel-alignment advantage of the analytic-on-analytic truth
construction; mechanism~3 carries the $3$--$7$~d horizons through
the integrated drag-acceleration error. The per-cell win
fractions of Table~\ref{tab:propagator-wins} match this
partition, including the $60.2\%$~$[55.9,\,64.4]$ hi-fid-wins
fraction at $550$~km~$\times$~$7$~d --- the cohort where SGP4's
own fit fails most visibly (super-linear $\hat k = 1.44$ on
v2-mini, Table~\ref{tab:powerlaw}) and where mechanism~2's
architectural advantage erodes against mechanism~3's drag-modelling
penalty.

\subsection{Cohort-resolved view: when does high-fidelity overtake SGP4?}
\label{sec:discussion-v2mini}

The shell-level wins fractions of Table~\ref{tab:propagator-wins}
report the H2 result as a single cohort-uniform exception at
$550$~km $\times$ $7$~d. As Section~\ref{sec:results-h2} flagged,
the per-(shell $\times$ $\Delta t$ $\times$ generation)
decomposition tells a more consistent story.
Table~\ref{tab:propagator-wins-by-gen} resolves every cell of the
corpus by pooled generation; three readings carry the H2-cohort
story.

\begin{table}[ht]
  \centering
  \caption{Per-(shell $\times$ $\Delta t$ $\times$ pooled
  generation) hi-fid-vs-SGP4 win fractions on 3D L$_{2}$ and on
  the along-track component, with $95\%$ satellite-level
  bootstrap percentile CIs. Same construction as
  Table~\ref{tab:propagator-wins} but with the cohort axis
  exposed; cells with fewer than $30$ pairs are suppressed. The
  v2-mini cohort at $\Delta t = 7$~d clears $50\%$ at both
  populated shells, in contrast with the shell-level reading in
  Table~\ref{tab:propagator-wins}.}
  \label{tab:propagator-wins-by-gen}
  \input{tables/tab_propagator_wins_by_gen.tex}
\end{table}

\emph{Cohort, not shell, carries the H2 sign at long $\Delta t$.}
On the v2-mini cohort alone, the high-fidelity arm wins
$72.6\%$~$[69.9,\,75.4]$ of $7$~d pairs at $550$~km and
$55.1\%$~$[49.9,\,60.4]$ of $7$~d pairs at $560$~km --- both
above the $50\%$ null. On the v1.x cohort alone the same
horizon reports $16.1\%$~$[11.0,\,22.2]$ at $550$~km and
$11.9\%$~$[8.8,\,15.5]$ at $560$~km --- both far below
$50\%$. The shell-level wins fractions in
Table~\ref{tab:propagator-wins} mix these two cohorts in
different population ratios (v2-mini is $78\%$ of the $550$~km
$\times$ $7$~d cell but only $34\%$ of the $560$~km $\times$
$7$~d cell), and the apparent shell-fragility of the H2-cohort
exception in the headline table is mostly that
blending artefact, not a cohort-level inconsistency. The same
table shows the v2-mini hi-fid-wins fraction climbing
monotonically with $\Delta t$ once past the $6$~h
truth-floor regime: from $\sim 17$--$45\%$ at $6$~h, to
$\sim 31\%$ at $1$~d, to $\sim 44$--$54\%$ at $3$~d, to
$\sim 55$--$73\%$ at $7$~d. The shell ordering is preserved
($550 > 560$) at long $\Delta t$, consistent with the lower-altitude
shell carrying the larger mean drag-induced
B$^{\!\star}$-mis-fit.

\emph{Median-level reading from Section~\ref{sec:results-h1}.}
The shell-level median table in
Section~\ref{sec:results-h1} already exposes the same cohort
ordering. At $\Delta t = 7$~d the v2-mini cohort medians sit
\emph{below} the SGP4 medians at both populated shells
(hi-fid $72.3$~km vs.\ SGP4 $101.7$~km on the $550$~km cell;
hi-fid $84.5$~km vs.\ SGP4 $89.8$~km on the $560$~km cell), while
on v1.x the hi-fid median sits a factor of three to five above
the SGP4 median at the same horizon
(Section~\ref{sec:results-h1}). The win-fraction picture and the
median picture therefore agree: hi-fid overtakes SGP4 on
v2-mini at long $\Delta t$ and loses to SGP4 on v1.x at the
same horizons; both signals are robustly present at both
populated shells.

\emph{Physical mechanism.} The cohort split has a coherent
physical reading that builds on mechanism 3 of
Section~\ref{sec:discussion-h2-mechanisms}. The v1.x bus has
been on-orbit since 2019--2021, with a multi-year operator
orbit-determination heritage; the fitted $B^{\!\star}$ on
v1.x absorbs the satellite's drag environment over its OD
window faithfully enough that SGP4's analytic propagation
beats a high-fidelity integrator whose drag inputs are derived
from public spacecraft properties. The v2-mini bus is newer
(2023+), with less-mature operator OD over an inherently
larger and more time-variable
$B^{\!\star}$ (the
$\hat k = 1.44$ super-linear SGP4 exponent on v2-mini at
$550$~km is the empirical signature of a constant unmodelled
along-track acceleration $a_{\rm{res}}$ that the operator-fitted
$B^{\!\star}$ has not absorbed; cf.\
Section~\ref{sec:methods-powerlaw-physics},
Eq.~\eqref{eq:powerlaw-k2}). On the same cohort, the
high-fidelity arm models drag explicitly through
$\rho\,C_{D}\,A$ from NRLMSISE-00 and the per-generation
spacecraft-property table, so where SGP4's averaged
$B^{\!\star}$ fails most badly the high-fidelity arm has the
most opportunity to gain. Lower altitude amplifies the effect:
more drag $\Rightarrow$ larger SGP4 mean-fit residual
$\Rightarrow$ more room for an explicit-drag propagator to
overtake. This is why the v2-mini wins fraction is higher at
$550$~km ($72.6\%$) than at $560$~km ($55.1\%$). The corpus
has no v2-mini cohort at $540$~km
(Table~\ref{tab:corpus}), so the lowest-altitude extrapolation
of the trend is untested.

\emph{Caveats.} Three constraints on the cohort claim should be
read alongside it. (i) The v2-mini drag area is scaled by the
v1.x-to-v2-mini bus-size ratio (Section~\ref{sec:methods-attitude},
Table~\ref{tab:drag-areas}), not fitted to v2-mini-specific
ballistic data --- the largest unconstrained modelling
assumption in the spacecraft-property layer. The
$\pm 20\%$ $C_{D} A$ sensitivity in
Appendix~\ref{app:cda-sensitivity} shifts the v2-mini median
high-fidelity error by at most $\pm 17\%$, leaving the
qualitative cohort sign intact, but a systematic v2-mini
$C_{D} A$ misspecification beyond that band would shift the
per-cohort wins fractions correspondingly. (ii) The
operator-fitted $B^{\!\star}$ on the newer cohort may carry
correlated biases (recent-maneuver history, attitude duty cycle
captured by the OD window) that this analysis cannot
separate from the H2 cohort signal using public TLEs alone.
(iii) The $560$~km $\times$ v2-mini $\times$ $7$~d cell carries
$n = 461$ pairs --- the smallest cell whose interpretation is
load-bearing for the cohort-uniform claim. A halving of the
corpus could move that cell's win fraction across the $50\%$
null, although the median ordering would survive at the
several-km gap reported in Section~\ref{sec:results-h1}.

\subsection{Along-track dominance of the error budget}
\label{sec:discussion-along-track}

Section~\ref{sec:results-error-structure} reports the empirical
signature: the along-track median sits a clean two decades above
cross-track and radial at every $\Delta t$ bucket of the corpus,
and by $\Delta t = 7$~d the L$_{2}$ norm and the along-track
component are indistinguishable within the line widths of
Figure~\ref{fig:error-decomposition}. The dominance is a textbook
consequence of orbital-dynamics geometry rather than a property of
either propagator. An in-track perturbation --- atmospheric drag,
an unmodelled component of the geopotential along the velocity
direction, or the constant in-track acceleration $a_{\rm{res}}$ of
Equation~\eqref{eq:powerlaw-k2} --- changes orbital energy, which
maps through the vis-viva relation to a mean-motion error
$\delta n$, which integrates over $\Delta t$ as the along-track
arc length $\Delta s \approx a\,\delta n\,\Delta t$ of
Equation~\eqref{eq:powerlaw-k1} (with $a_{\rm{res}}$-driven
contributions adding the quadratic-in-$\Delta t$ term). Radial
and cross-track errors, by contrast, are bounded under the same
perturbations: $J_{2}$ nodal regression and eccentricity-vector
drift oscillate at the orbital period and grow only secularly
through second-order coupling, so neither component accumulates
freely with $\Delta t$. This is the canonical ``along-track
linear, transverse bounded'' signature reported in every LEO
position-error study from \citet{vallado2006} onward, formalised
through the Hill / Clohessy--Wiltshire relative-motion equations
\citep[\S\,3.4]{vallado2013}, and codified in operational SSA
where conjunction-data messages and orbit-determination
residuals quote the along-track component separately as a matter
of course.

The H2 mechanism partition of
Section~\ref{sec:discussion-h2-mechanisms} therefore lives
predominantly on the along-track axis. At-epoch operator OD
residuals and $B^{\!\star}$ mis-fits both project most strongly
onto the velocity-aligned component --- both are absorbed by the
operator into a single scalar at $t_{i}$, and that scalar's
mismatch against the forward propagation environment grows
linearly along-track. Empirically, the along-track-only variant
of the per-pair win-fraction
$\lvert\Delta r_{\mathrm{along},\,\mathrm{hifi}}\rvert <
\lvert\Delta r_{\mathrm{along},\,\mathrm{SGP4}}\rvert$
(Table~\ref{tab:propagator-wins}) moves the $6$~h hi-fid-wins
fraction from $33.7\%$ to $36.0\%$ but is essentially unchanged
at $1$~d ($25.0\%$), $3$~d ($28.7\%$), and $7$~d ($35.2\%$);
the L$_{2}$ comparison is already an along-track comparison once
the orbital geometry pushes the radial and cross-track
contributions two decades below. The H2 negative result is
robust to the choice of metric.

The per-cell $(A,\,k)$ table summarised by
Figure~\ref{fig:powerlaw-fits} and tabulated in
Table~\ref{tab:powerlaw} can serve a complementary downstream use
beyond the present comparison, as a benchmark target for the
enhanced-propagator line of work surveyed in
Section~\ref{sec:background}: SGP4-XP, differentiable SGP4, and
ML-residual correctors all retain the SGP4 evaluation budget and
the TLE input format, so a future enhanced propagator measured on
the same locked corpus and $\Delta t$ buckets, with $(A,\,k)$
refitted under the same methodology, would yield a directly
comparable per-cell pair. The atlas is in a form usable as a
regression test by that work, independently of whether the H2
sign reported here survives propagator innovation.

\subsection{Practical implications for public-TLE consumers}
\label{sec:discussion-practical}

Starting from public TLEs, a GMAT-class high-fidelity propagator
configured per Table~\ref{tab:force-model} (EGM2008
$70\!\times\!70$, Sun and Moon point-mass third bodies,
NRLMSISE-00 drag, conical-shadow SRP) and supplied with the
per-generation drag- and SRP-area estimates of
Section~\ref{sec:methods-spacecraft-props} does not improve median
3D position error over SGP4 at any of the four staleness horizons
sampled by this corpus. The pooled hi-fid-wins fraction sits at
$33.7\%$~$[32.2,\,35.3]$ at $\Delta t = 6$~h, dips to
$25.1\%$~$[23.9,\,26.1]$ at $1$~d, and recovers to
$35.3\%$~$[31.9,\,38.8]$ at $7$~d
(Table~\ref{tab:propagator-wins}); the median high-fidelity error
exceeds the SGP4 median by factors of roughly $1.6$, $2.0$, $2.1$,
and $2.0$ at the four buckets (Section~\ref{sec:results-h1}). The
qualitative pattern is robust to $\pm 20\%$ $C_{D} \cdot A$
uncertainty on the v2-mini cohort
(Appendix~\ref{app:cda-sensitivity}), to $\pm 50$~m perturbations
around the $100$~m maneuver-filter threshold
(Appendix~\ref{app:maneuver-threshold-sensitivity}), and to the
choice between the daily-mean and 81-day-centred F10.7
predictors in the H3 fit (Section~\ref{sec:methods-h3-regression}).

For practitioners the operational takeaway is asymmetric across
horizons. Inside one operator update interval of TLE publication
--- the $\sim 4$~h Starlink cadence, captured by the $6$~h bucket
of this corpus --- the propagator choice is secondary to the
mechanism-1 OD-residual floor: both arms cluster near $1$~km median
L$_{2}$ error against the proxy truth, with the inter-propagator
gap dominated by mechanism~2's kernel-alignment offset rather than
by physical-fidelity differences. At the $1$--$3$~d horizon a
high-fidelity workflow that brings independent ballistic-coefficient
estimates and an operator-grade initial state (e.g.\ from internal
OD or from a third-party tracking service) could in principle close
mechanism-3 contributions; the present result places the upper
bound on that closure when only public-TLE inputs are available.
At the $7$~d horizon the high-fidelity arm is behind SGP4 by a
factor approaching $3.4$, mechanism~3 dominates the differential,
and the improvement window for a public-data consumer is closed
unless the initial state itself improves.

The cohort-resolved analysis of
Section~\ref{sec:discussion-v2mini} identifies one cohort
(v2-mini) at one staleness horizon ($\Delta t \gtrsim 3$~d) where
high-fidelity propagation overtakes SGP4 on a majority of pairs
at \emph{both} populated shells: the v2-mini hi-fid-wins fraction
on $7$~d pairs is $72.6\%$~$[69.9,\,75.4]$ at $550$~km and
$55.1\%$~$[49.9,\,60.4]$ at $560$~km
(Table~\ref{tab:propagator-wins-by-gen}). The regime is the one
where SGP4 itself struggles most --- super-linear
$\hat k \approx 1.44$ on v2-mini at $550$~km and $1.32$ at
$560$~km, on a cohort whose operator-fitted $B^{\!\star}$ tracks
density less stably than v1.x at the same altitude --- so the
architectural advantage of mechanism~2 erodes against the
underlying skill gap, and mechanism~3's drag-modelling penalty
flips sign on the cohort where SGP4 is weakest. Consumers whose
workflow centres on long-staleness v2-mini forecasting should
regard the propagator choice as cohort-specific rather than
uniformly disfavouring high-fidelity. For the same reason the per-cell $(A,\,k)$ table
of Table~\ref{tab:powerlaw} is the natural input to a
practitioner's error budget: it preserves the cohort-resolved
$A$ and $k$ that no single pooled number captures.

\subsection{Covariance growth and conjunction screening}
\label{sec:discussion-covariance}

The present analysis reports deterministic L$_{2}$ norms and signed
RSW components of the position-error vector. A complete uncertainty
decomposition of the propagation pipeline --- initial-state
covariance mapped forward by the state-transition matrix,
force-model parameter uncertainty, density-model uncertainty, the
operator-OD-residual distribution at both endpoints --- is left to
a follow-up study; the deterministic staleness atlas provided here
is the \emph{baseline curve} that any subsequent Monte Carlo or
sigma-point uncertainty study would integrate against, and the
per-cell $(A,\,k)$ parametrisation of Table~\ref{tab:powerlaw} is
the natural reference for a probabilistic extension that fits the
same functional form to the propagated covariance trace.

For conjunction-screening practitioners, the along-track-dominated
character of the error
(Sections~\ref{sec:results-error-structure} and
\ref{sec:discussion-along-track}) maps directly to
timing-residual covariance: the $1$~km / $0.13$~s
correspondence at the $550$~km shell sets the scale at which a
deterministic miss-distance becomes a probabilistic one through
the covariance ellipse. The deterministic divergence
characterised here is \emph{not} the covariance growth of the
propagator --- those are formally distinct quantities, the former
a single trajectory's drift against a proxy truth and the latter
the trace of the propagated covariance matrix --- but the present
results constrain a useful aspect of the latter: any covariance
realism study must reproduce the per-cell deterministic curves of
Table~\ref{tab:powerlaw} as a special case at zero process noise,
since the deterministic and stochastic propagations share the same
underlying force model and the same initial-state mean. Operational conjunction-data
messages quote along-track $1\sigma$ separately from the radial
and cross-track $1\sigma$ precisely because that decomposition
mirrors the error budget the present analysis isolates.

\subsection{Limitations}
\label{sec:discussion-limitations}

The analysis is constrained on several axes worth listing
explicitly. \emph{Single constellation.} Starlink dominates the
public-TLE LEO population but is not the only platform of
interest; OneWeb, Kuiper, and the small-satellite fleets carry
different bus geometries and ballistic-coefficient scales and are
out of scope here. The three-mechanism partition of
Section~\ref{sec:discussion-h2-mechanisms} is constellation-agnostic
on its mechanism descriptions but the quantitative win-fraction
levels are not.

\emph{30-day window of moderate solar activity.} The corpus spans
2026-04-01 to 2026-05-01 inclusive, with a daily-mean
F10.7 range of $\sim 17$~sfu across satellites and an
81-day-centred range of $\sim 1.3$~sfu. The H3 fit's leverage on
the long-period predictor is correspondingly small
(Section~\ref{sec:results-h3}); extrapolation to high-F10.7
conditions, where atmospheric density at $540$~km can be a factor
of several over the corpus median, inherits the assumption that
the present mechanism partition holds.

\emph{Atmosphere model.} NRLMSISE-00 is the SSA community
standard at LEO altitudes today, but its successor NRLMSIS~2.0/2.1
\citep{emmert2021} can shift modelled densities by $10$--$20\%$
over the present model in the F-region thermosphere, and
operational space-surveillance systems use JB2008
\citep{bowman2008}, which GMAT~R2026a does not bundle. A
sensitivity rerun under either alternative is the most direct test
of the mechanism-3 calibration assumption.

\emph{Truth construction.} The next-TLE-as-proxy methodology is by
assumption a self-consistency check against the operator's own
orbit determination
(Section~\ref{sec:background-truth-floor}); the
$\Delta\mathbf{r}_{\mathrm{OD}}$ residuals at both endpoints are
real and bound the absolute interpretability of the $6$~h headline
in particular. A true ground-truth assessment would require
operator-internal GNSS or radar-tracked state vectors, neither of
which is currently available for public study.

\emph{Spacecraft-property prior on v2-mini.} The v2-mini effective
drag area at $C_{D} = 2.2$ is scaled from the Baruah v1.x
shark-fin anchor by the published bus-size ratio
(Section~\ref{sec:methods-attitude}, Table~\ref{tab:drag-areas});
Appendix~\ref{app:cda-sensitivity} shows the H2 reading robust to
$\pm 20\%$, but absolute v2-mini error magnitudes carry that
calibration uncertainty as a common-mode offset.

\emph{Maneuver-filter sensitivity floor.} Continuous low-thrust
station-keeping is detected only through its per-step SMA jumps;
a hypothetical purely continuous-drift mode whose per-step
$\lvert\Delta a\rvert$ never crosses the calibrated threshold
slips through the filter (Section~\ref{sec:methods-pairs},
\citealp[\S\,3.2]{lemmens2014}).
Appendix~\ref{app:maneuver-threshold-sensitivity} bounds the
effect of the threshold choice in the empirically interesting
$50$--$200$~m region, but the limit-of-detection floor remains
methodological.

\subsection{Extensions}
\label{sec:discussion-extensions}

Three concrete extensions follow naturally from the present
analysis. The most informative is replication on a longer window
straddling a solar-cycle transition: the narrow F10.7 lever on the
30-day corpus is the principal limitation on the H3 statistical
power (Section~\ref{sec:results-h3}), and a $6$--$12$-month corpus
spanning both quiet and active periods would yield the cleanest
test of the density-modulation hypothesis. The atmosphere-model
swap to NRLMSIS~2.1 \citep{emmert2021} or JB2008
\citep{bowman2008} is the second: the per-cell $(A,\,k)$ table is
a ready-made benchmark against which an atmosphere-model
sensitivity study would shift the high-fidelity arm in a directly
comparable way. The third is cross-constellation replication on
OneWeb, Kuiper, or the small-satellite fleets with the same
methodology and the same locked corpus structure, both to test
the cohort universality of the three-mechanism partition and to
provide a cross-constellation reference for conjunction-assessment
practitioners.

Two methodological extensions are also in scope. The
mixed-effects supplement of Appendix~\ref{app:mixed-effects}
reports per-cell linear mixed-effects fits as a parametric
cross-check of the per-pair OLS estimator with satellite-level
bootstrap CIs; in a follow-up release the mixed-effects fit
could be escalated to the main-body estimator if the per-cell
agreement of Table~\ref{tab:mixed-effects} held on extended
corpora, or if a reviewer pushed for the parametric absorption
of within-sat correlation rather than the present non-parametric
resample. A gravity-truncation sensitivity test at
$40\!\times\!40$ and $20\!\times\!20$ would quantify the (small)
effect of higher-order geopotential terms on the long-$\Delta t$
high-fidelity arm and is straightforward to add against the
existing sweep harness; the test is one of methodological
completeness rather than expected sensitivity, given that the
EGM2008 truncation error at $70\!\times\!70$ at LEO altitudes
\citep{pavlis2012} sits well below the drag- and
OD-residual-driven kilometre-scale errors the high-fidelity arm
already carries.

A complete uncertainty-propagation pipeline --- propagating the
operator-OD covariance forward and combining it with force-model
and atmosphere-model parameter uncertainty --- is the natural
follow-up of largest interest to conjunction-screening
practitioners and is the principal direction the deterministic
atlas reported here is intended to enable. The per-cell
$(A,\,k)$ pair is the form that a probabilistic extension would
read against, and the architecture and reproducibility surface
of the present analysis
(Section~\ref{sec:methods-reproducibility}) are designed so the
extension can be driven by a re-run of the same sweep with a
covariance-aware integrator and an unchanged corpus.

\section{Conclusion}
\label{sec:conclusion}

We have characterised the position-error behaviour of SGP4 and a
GMAT-class high-fidelity propagator against operator-updated truth on
the Starlink megaconstellation, sweeping $24{,}641$ next-TLE-truth
pairs across $501$ satellites stratified by altitude shell and platform
generation. The three hypotheses introduced in
Section~\ref{sec:introduction} resolve as follows:
each populated (altitude shell $\times$ pooled generation) cell
of the corpus carries a statistically distinguishable power-law
staleness curve $\lVert\Delta\mathbf{r}(\Delta t)\rVert \approx
A\,\Delta t^{k}$, with fitted exponents in $(1,\,2)$ on every
v2-mini cell and on the high-fidelity v1.x cells at $540$ and
$560$~km, and sub-linear SGP4 v1.x cells plus an at-boundary
high-fidelity v1.x cell at $550$~km that the
$k=1$ likelihood-ratio test does not reject (H1,
Table~\ref{tab:powerlaw}); the
high-fidelity propagator initialised from a public TLE does
\emph{not} improve median position error over SGP4 at any of the
four shell-pooled staleness horizons sampled by this corpus, losing
on $\sim 65$--$75\%$ of pairs depending on $\Delta t$, with the
exception of the v2-mini cohort at long $\Delta t$ where
high-fidelity overtakes SGP4 on a majority of pairs at both
populated shells (H2, Tables~\ref{tab:propagator-wins} and
\ref{tab:propagator-wins-by-gen},
Section~\ref{sec:discussion-v2mini}); and an exploratory regression of
the per-satellite SGP4 staleness coefficient $A$ against F10.7
returns a slope that is direction-consistent with the
$\partial\,\log\rho / \partial\,\mathrm{F}10.7 > 0$ expectation at
LEO but clears conventional significance at only one of the three
shells ($560$~km) on this $30$-day, moderate-activity,
$\sim 17$~sfu-span corpus, and should be read as preliminary
rather than as a calibrated density-modulation measurement (H3,
Figure~\ref{fig:solar-modulation}).

Three practitioner-facing summaries condense the operational reading
of these results. First, Starlink TLEs may be used unmodified up to
the operator's $\sim 4$-hour refresh interval; consumers should
expect a $\sim 1$~km median 3D position error at the $6$~h boundary
across all shells and generations, with the bulk of that error
attributable to the operator's at-epoch OD residual rather than to
propagator quality
(Section~\ref{sec:discussion-h2-mechanisms}, mechanism 1). Second,
high-fidelity propagation should not be invested in for public-TLE
workflows unless the consumer also brings better-than-operator-OD
initial states or operates on the v2-mini cohort at long
$\Delta t$; the gain over SGP4 from a force-model upgrade alone is
negative or marginal in every shell-pooled cell of the corpus
except the cohort-resolved v2-mini long-$\Delta t$ regime, and the
asymmetric mechanism-3 amplification of spacecraft-property
uncertainty on the high-fidelity arm sets an upper bound on what a
$\pm 20\%$ improvement in $C_{D} \cdot A$ knowledge can recover
(Appendix~\ref{app:cda-sensitivity}). Third, the v2-mini cohort is
the high-uncertainty cohort \emph{and} the cohort where the
propagator choice is most defensible: SGP4 $7$~d medians are
$2$--$4\times$ higher than v1.x at the same altitude
(Section~\ref{sec:results-h1}), operational staleness budgets on
that cohort should be sized $\sim 2\times$ looser than v1.x
correspondingly, and at $\Delta t \gtrsim 3$~d the high-fidelity
arm becomes the better predictor on a majority of v2-mini pairs at
both $550$ and $560$~km
(Section~\ref{sec:discussion-v2mini},
Table~\ref{tab:propagator-wins-by-gen}).

Beyond these three takeaways, the per-cell power-law atlas of
Table~\ref{tab:powerlaw} is intended to be \emph{reusable} as a
benchmark target for the enhanced-propagator line of work surveyed
in Section~\ref{sec:background}: the same locked corpus, the same
$\Delta t$ buckets, and the same $(A,\,k)$ estimator are available
for any future propagator to refit against, and the cohort-resolved
table is the natural form for a downstream error-budget input. The
deterministic baseline reported here provides the curve a Monte
Carlo or sigma-point uncertainty study would integrate against
(Section~\ref{sec:discussion-covariance}). All code, the locked
corpus, the resumable sweep manifest, and the rendered figures are
released together under the MIT license at the URL listed in
Section~\ref{sec:availability}.

\appendix

\section{Maneuver filter calibration}
\label{app:maneuver-filter}

\begin{figure}[ht]
  \centering
  \includegraphics[width=\linewidth]{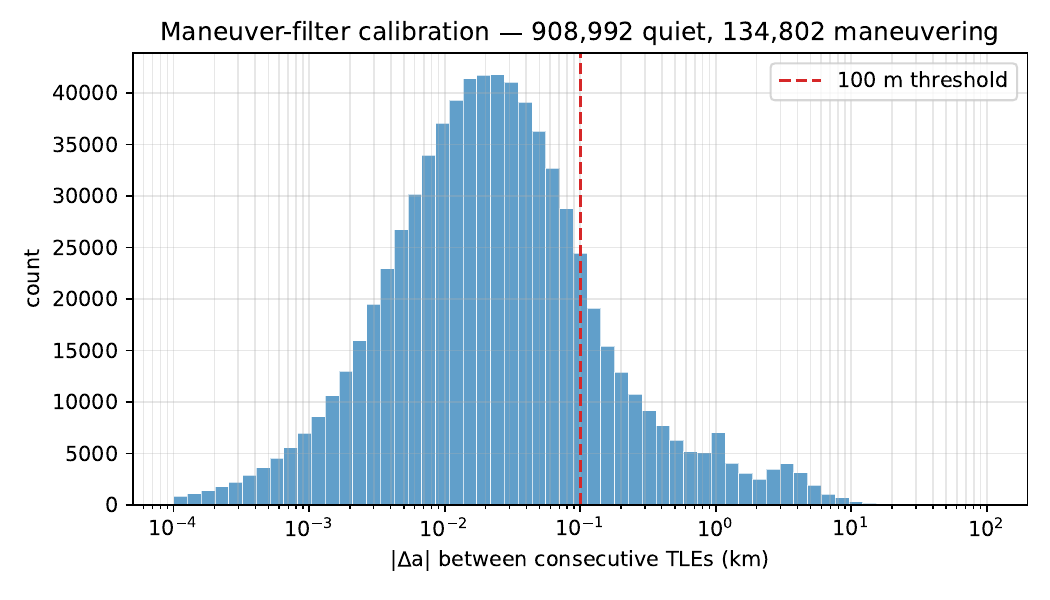}
  \caption{Distribution of $|\Delta a|$ between consecutive Starlink
  TLEs across the raw cache. The dashed line marks the 100\,m maneuver
  filter threshold. The OD-noise mode and the maneuver mode overlap in
  the $50$--$200$~m region rather than separating at a clean valley
  (the expected signature of Starlink's continuous low-thrust
  station-keeping cadence); the 100\,m threshold cuts at the right
  edge of the OD-noise mode, and the
  Appendix~\ref{app:maneuver-threshold-sensitivity} sensitivity
  perturbation tests the choice at $\pm 50$~m.}
  \label{fig:maneuver-filter}
  \script{fig_maneuver_filter.py}
\end{figure}

The pair-construction pipeline drops consecutive TLE pairs whose
mean-motion derived semi-major-axis difference $|\Delta a|$ exceeds a
fixed threshold, on the grounds that such jumps mark station-keeping
burns and so violate the no-maneuver assumption behind the truth
construction. Figure~\ref{fig:maneuver-filter} shows the
$|\Delta a|$ distribution across all consecutive pairs in the raw
cache, with the 100\,m threshold marked.

The per-cell rejection footprint of the filter is summarised in
Table~\ref{tab:maneuver-rejections}: at the baseline 100~m threshold,
the filter rejects $5{,}810$ of the $30{,}451$ candidate pairs
($19.1\%$) across the 12 populated $(\text{altitude shell} \times
\Delta t)$ cells. The rejection rate climbs monotonically with
$\Delta t$ within each shell, as the half-open interval
$(t_{i},\,t_{j}]$ has more opportunity to enclose a station-keeping
event, and with shell altitude, consistent with the v2-mini cohort's
higher operational maneuver cadence relative to v1.5 at lower shells.
The $24{,}641$ survivors are the production corpus used everywhere
else in this paper.

\begin{table}[ht]
  \centering
  \caption{``Fleet quietness'' summary at the baseline 100~m maneuver
  threshold. Counts are aggregated across the 501-satellite stratified
  sample (Section~\ref{sec:methods-pairs}) into the
  $(\text{altitude shell} \times \Delta t)$ cells underlying
  Table~\ref{tab:corpus}; the rejection fraction is the share of
  candidate pairs whose $(t_{i},\,t_{j}]$ interval covers an SMA-jump
  event exceeding the threshold.}
  \label{tab:maneuver-rejections}
  \input{tables/tab_maneuver_rejections.tex}
\end{table}

\section{Hexbin density view of the propagator scatter}
\label{app:propagator-scatter-hexbin}

\begin{figure}[ht]
  \centering
  \includegraphics[width=\linewidth]{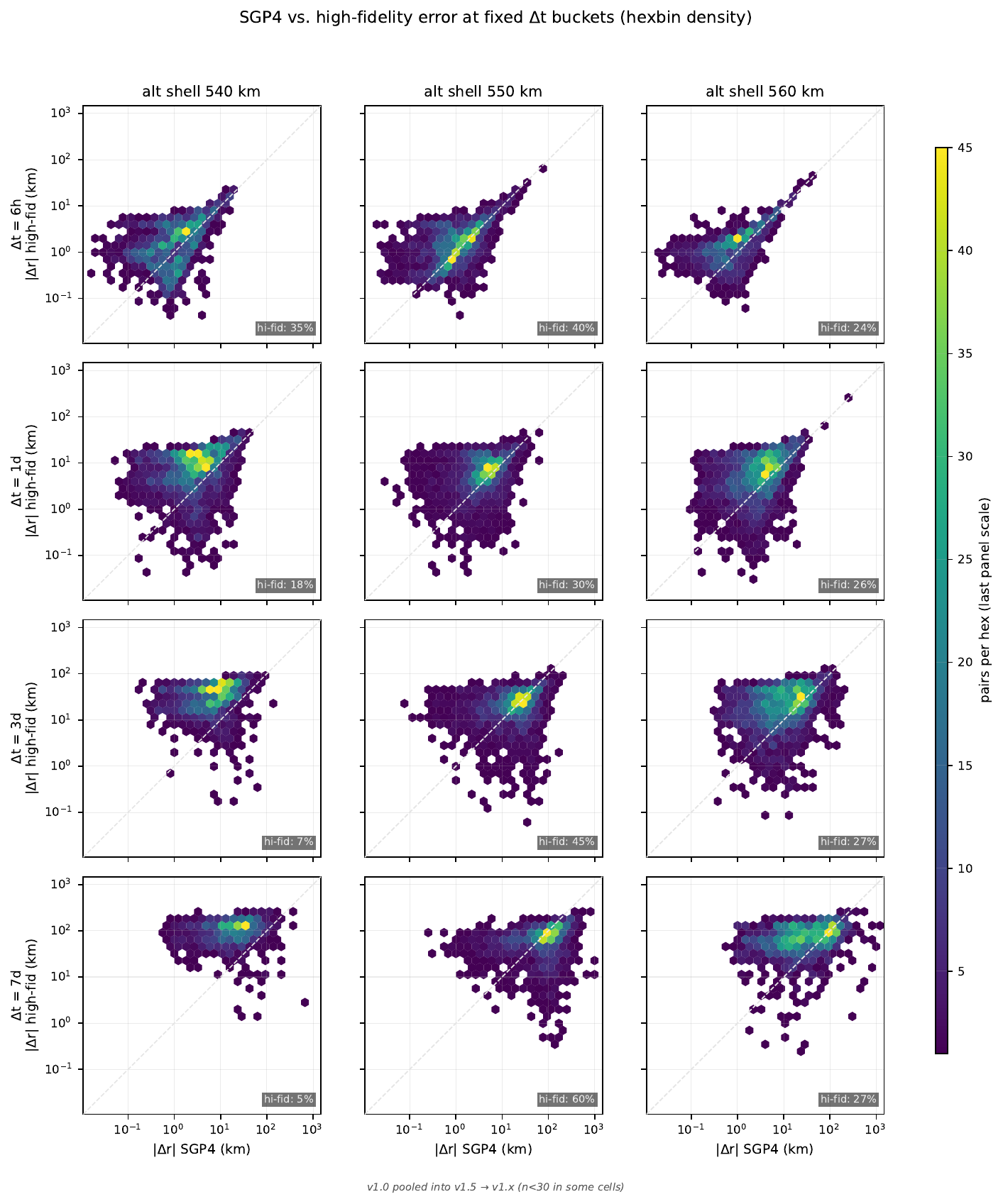}
  \caption{Hexbin density variant of
  Figure~\ref{fig:propagator-scatter}. Each panel is binned in
  $(\log_{10}|\Delta\mathbf{r}|_{\mathrm{SGP4}},\,\log_{10}|\Delta
  \mathbf{r}|_{\mathrm{hifi}})$ on $\mathrm{gridsize} = 30$ hexes
  per side, viridis colour ramp with $\mathrm{mincnt} = 1$;
  per-panel hi-fid-wins fraction inset as in the main scatter.}
  \label{fig:propagator-scatter-hexbin}
  \script{fig_propagator_scatter_hexbin.py}
\end{figure}

Figure~\ref{fig:propagator-scatter} in the main body shows every
$(\mathrm{TLE}_{i},\,\mathrm{TLE}_{j})$ pair as an individual marker
on the SGP4-vs-high-fidelity plane, so that points above and below
the $y = x$ diagonal can be read pair by pair. At the longer
$\Delta t$ buckets the per-pair markers begin to overplot, masking
the joint-density structure of the comparison.
Figure~\ref{fig:propagator-scatter-hexbin} renders the same data on
the same axes as a log-binned hexbin density. The horizontal
saturation band at $540$~km~$\times$~$\{3,\,7\}$~d --- the
fingerprint of a near-constant high-fidelity force-model floor ---
becomes a single sharp ridge of the densest hex cells, and the
v2-mini wedge of below-the-diagonal pairs at
$\{550,\,560\}$~km~$\times$~$7$~d is visible as a low-density
plume reaching well into the high-fidelity-wins region. The
per-panel hi-fid-wins fraction inset is repeated here so the
density view and the scatter view share an identical numerical
summary.

\section{Drag-area sensitivity (v2-mini)}
\label{app:cda-sensitivity}

Section~\ref{sec:methods-attitude} adopts the shark-fin attitude as
the duty-cycle-representative drag mode, anchored to
\citet{baruah2024} for v1.x and bus-scaled for v2-mini --- the
latter being the largest unconstrained modelling assumption in the
spacecraft-properties layer. We probe its impact on the H2 result
by re-running the v2-mini slice of the stratified 1{,}000-pair
sensitivity subset (Section~\ref{sec:methods-pairs}) at $C_D \cdot
A$ scaling factors of $0.8$ and $1.2$, holding every other input
(initial state, dry mass, $C_D$, SRP area, atmosphere model,
integrator settings, space weather) constant. The baseline at
factor $1.0$ is the same v2-mini slice of the main-sweep aggregate
\texttt{outputs/all\_runs.parquet}; the two off-baseline frames are
\texttt{outputs/all\_runs\_cda\_low.parquet} and
\texttt{outputs/all\_runs\_cda\_high.parquet} respectively, produced
by the \texttt{sweep.cda\_sensitivity} driver
(\texttt{make cda-sensitivity}).

\begin{table}[ht]
  \centering
  \caption{Median high-fidelity position error
  $|\Delta\mathbf{r}|_{\mathrm{hifi}}$ for the v2-mini cohort by
  altitude shell and $\Delta t$ bucket, at the three $C_D \cdot A$
  scaling factors $\{0.8,\,1.0,\,1.2\}$ relative to the
  Section~\ref{sec:methods-attitude} baseline.
  The $0.8\times$ and $1.2\times$ columns report the median expressed as a
  relative shift versus the baseline. The 540~km shell is absent because
  the corpus has no v2-mini satellites at that altitude
  (Table~\ref{tab:corpus}).}
  \label{tab:cda-sensitivity}
  \input{tables/tab_cda_sensitivity.tex}
\end{table}

Across the eight populated $(\mathrm{shell},\,\Delta t)$ cells the
$\pm 20\%$ $C_D \cdot A$ perturbation shifts the median
$|\Delta\mathbf{r}|_{\mathrm{hifi}}$ by between $\pm 1.4\%$ (at
$\Delta t = 6$~h, where the OD-residual floor dominates) and
$\pm 17.1\%$ (at $\Delta t = 7$~d, where drag has integrated long
enough for the $C_D \cdot A$ bias to express itself); every cell
moves in the expected direction (larger drag area
$\Rightarrow$ larger hi-fid error). Pooled across cells, the median
shifts $-8.9\%$ at factor $0.8$ and $+11.0\%$ at factor $1.2$. The
maximum absolute shift sits well inside the SGP4-vs-hi-fid gap
quantified in Section~\ref{sec:discussion}
($+240\%$ at the same $7$~d horizon), so the H2 negative result
survives intact under realistic in-attitude $C_D \cdot A$
uncertainty on the v2-mini cohort. The increasing shift magnitude
with $\Delta t$ is the integrated-drag signature expected on
physical grounds: at short horizons the OD-residual floor masks
the drag perturbation; at longer horizons the perturbation
accumulates and the SGP4 arm absorbs the equivalent bias into its
operator-fitted $B^{\!\star}$, while the hi-fid arm carries it as
a propagated error.

\section{Maneuver-threshold sensitivity}
\label{app:maneuver-threshold-sensitivity}

Section~\ref{sec:methods-pairs} adopts a 100~m SMA-jump threshold for
the maneuver filter, calibrated empirically against the $|\Delta a|$
histogram of Figure~\ref{fig:maneuver-filter}. The OD-noise mode and
the maneuver mode overlap in the $50$--$200$~m region, so a
$\pm 50$~m perturbation around the baseline threshold could in
principle shift the locked sweep corpus in either direction. We
probe the effect of that perturbation on the headline medians by
rebuilding the candidate corpus at the two threshold endpoints and
comparing per-cell median high-fidelity position errors against the
baseline. The 50~m corpus is a strict subset of the 100~m corpus
($20{,}639$ surviving pairs vs.\ $24{,}641$) and so reuses the
existing GMAT outputs filtered to the surviving pair set; the 200~m
corpus adds $2{,}449$ pairs that the 100~m filter rejected
(``augment'' pairs, $27{,}090$ total at 200~m), and we re-run the
high-fidelity sweep on those pairs to obtain a comparable per-cell
median. The augment driver
(\texttt{sweep.maneuver\_threshold\_sensitivity},
\texttt{make maneuver-threshold-sensitivity}) and the table emitter
(\texttt{sweep.maneuver\_threshold\_table},
\texttt{make maneuver-threshold-table}) produce
Table~\ref{tab:maneuver-threshold-sensitivity} and the corresponding
\texttt{outputs/maneuver\_threshold\_summary.json}; both reuse the
RunSpec / preprocessing helpers from \texttt{sweep.run\_sweep} so the
augment arm is bit-comparable to the main hi-fid sweep.

\begin{table}[ht]
  \centering
  \caption{Per-cell median high-fidelity position error
  $|\Delta\mathbf{r}|_{\mathrm{hifi}}$ at the baseline 100~m maneuver
  threshold, with its $95\%$ satellite-level bootstrap confidence
  interval (1{,}000 resamples drawing satellites with replacement and
  pooling all of each drawn satellite's pairs to preserve within-sat
  correlation across $\Delta t$ buckets). The 50~m and 200~m columns
  report the per-cell median expressed as a relative shift versus the
  baseline; values flagged with $\dagger$ sit outside the $95\%$
  bootstrap CI of the corresponding baseline cell.}
  \label{tab:maneuver-threshold-sensitivity}
  \input{tables/tab_maneuver_threshold.tex}
\end{table}

Across the 12 populated $(\mathrm{shell},\,\Delta t)$ cells the per-cell
median high-fidelity position error moves by at most $-5.7\%$ at the
$50$~m threshold (worst-cell $550$~km $\times$ $6$~h) and at most
$+3.4\%$ at the $200$~m threshold ($560$~km $\times$ $6$~h);
$11$ of the $12$ cells have a $50$~m median inside the corresponding
baseline $95\%$ bootstrap confidence interval, and all $12$ have a
$200$~m median inside it. The single cell whose $50$~m median lands
outside its baseline CI is $540$~km $\times$ $7$~d, where the
baseline CI of $[166.0,\,180.4]$~km is unusually tight ($\sim 8.5\%$
wide) and a $-4.4\%$ shift in the median to $165.6$~km slips just
below the lower bound; the perturbation magnitude is comparable to
the other cells but the discriminating power of the CI is larger at
that horizon. Every cell moves in the expected direction (a stricter
threshold rejects pairs adjacent to maneuvers, biasing the surviving
distribution toward smaller errors, and conversely for a looser
threshold). The maximum per-cell shift is an order of magnitude
smaller than the $\pm 17\%$ peak observed under the
$\pm 20\%$ $C_D \cdot A$ perturbation
(Appendix~\ref{app:cda-sensitivity}) and two orders of magnitude
below the SGP4-vs-hi-fid gap quantified in
Section~\ref{sec:discussion}, so the H2 negative result is robust
against the threshold-overlap region flagged by reviewers as a
methodological concern.

The 50~m corpus retains $20{,}639$ pairs, the baseline $100$~m corpus
retains $24{,}641$, and the $200$~m corpus retains $27{,}090$
($2{,}449$ augment pairs whose $(t_{i},\,t_{j}]$ interval contains
an SMA-jump of magnitude $100$--$200$~m); the inclusion chain
$50\,\text{m} \subset 100\,\text{m} \subset 200\,\text{m}$ is verified
empirically at corpus-build time. The bootstrap CI is computed over
$1{,}000$ satellite-level resamples, drawing satellites with
replacement and pooling all of each drawn satellite's pairs in the
cell, so within-sat correlation across the four $\Delta t$ buckets
does not inflate the percentile interval.

\section{Pair-matching selection-effect diagnostic}
\label{app:selection-effect}

\begin{figure}[ht]
  \centering
  \includegraphics[width=\linewidth]{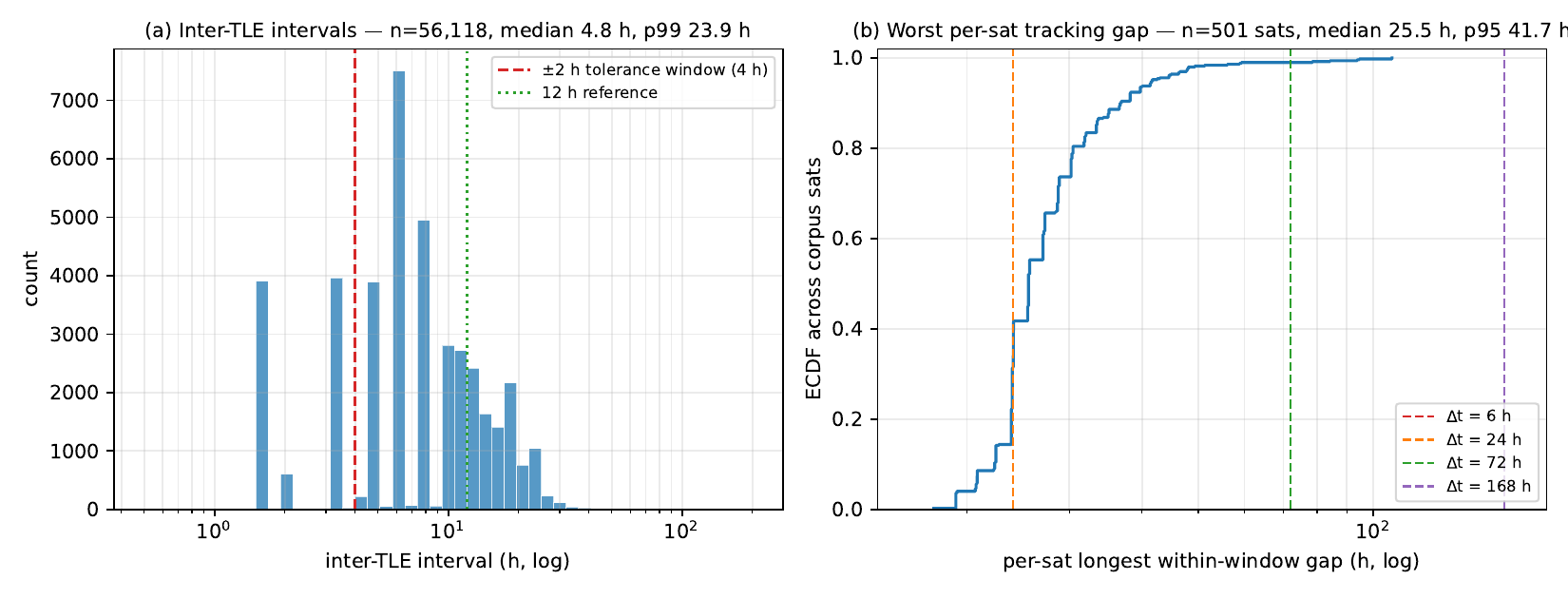}
  \caption{Selection-effect diagnostic for the $\pm 2$~h
  pair-matching tolerance. (a) Histogram of inter-TLE intervals
  across the corpus (log-x); the dashed line marks the
  $4$~h-wide tolerance window, the dotted line a $12$~h
  reference. (b) Empirical CDF of per-sat longest within-window
  gaps across the 501 sampled sats; vertical lines mark the four
  $\Delta t$ targets. The 95th-percentile worst per-sat gap
  ($\sim 42$~h) is well inside the longest $\Delta t$ target
  ($168$~h), so no sat is systematically excluded by the
  matching tolerance.}
  \label{fig:selection-effect}
  \script{fig_selection_effect.py}
\end{figure}

Section~\ref{sec:methods-pairs} quotes two distinct quantities to
support the claim that the $\pm 2$~h pair-matching tolerance does
not systematically bias the corpus against periods of poor
operator tracking. Figure~\ref{fig:selection-effect} visualises
both. Panel~(a) shows the distribution of inter-TLE intervals
across the 56{,}118 per-sat consecutive pairs in the 501-sat
corpus population; the bulk clusters near the $\sim 4.8$~h
median, with a right tail extending past 24~h, and the $\pm 2$~h
($4$~h-wide) matching window is the same order as the typical
inter-TLE spacing --- so a randomly-placed target window will
often contain zero TLEs by Poisson statistics alone, independent
of any tracking quality variation. Panel~(b) shows the empirical
cumulative distribution of the per-sat longest within-window gap.
The 95th-percentile worst per-sat gap sits at $\sim 42$~h, well
inside the longest $\Delta t$ target ($168$~h), so every
corpus satellite produces matched pairs at most of its
sat-days at the 1, 3, and 7~d horizons; only the 6~h target is
materially affected by the right tail, and only by a Poisson
mechanism that is roughly uniform across sats. Together these
panels are the empirical basis for the §\,\ref{sec:methods-pairs}
conclusion that the corpus is mildly biased toward well-tracked
sat-days but is not asymmetrically excluding any satellite.

\section{Mixed-effects supplement}
\label{app:mixed-effects}

Section~\ref{sec:methods-statistics} adopts a per-pair OLS power-law
fit with satellite-level bootstrap CIs as the main estimator for
the $(A,\,k)$ parameters reported in Table~\ref{tab:powerlaw}. The
satellite-level resample is the non-parametric concession to
within-sat correlation across the four $\Delta t$ buckets: it makes
no parametric assumption about how an individual satellite's pairs
covary, but it also produces no parameter for that covariance that
a reader could inspect. The parametric counterpart is a linear
mixed-effects (LME) fit of
\[
  \log_{10}\lVert\Delta\mathbf{r}\rVert \;\sim\;
    \log_{10}(\Delta t / 1\,\text{h}) \;+\; (1 \mid \texttt{norad\_id}),
\]
with the random intercept absorbing per-satellite shifts in
$\log_{10} A$ and the fixed-effect slope playing the role of the
power-law exponent $k$. We fit this model per
(altitude shell $\times$ pooled generation $\times$ propagator)
cell via REML with statsmodels' default BFGS optimiser
(\texttt{statsmodels.formula.api.mixedlm}). L-BFGS was tried
first but collapsed the random-intercept variance to the zero
boundary on the well-populated cells of this corpus and aliased
the scale into the fixed-effect intercept, so the default BFGS
optimiser is the load-bearing convergence choice here. The driver
\texttt{python src/scripts/\_mixed\_effects.py} emits the full
per-cell payload to
\texttt{outputs/mixed\_effects\_results.csv} and the booktabs
comparison fragment of Table~\ref{tab:mixed-effects} to
\texttt{src/tex/tables/tab\_mixed\_effects.tex}.

The supplement serves two purposes. First, the random-intercept
standard deviation on $\log_{10} A$ is a directly-interpretable
measure of the cross-satellite scatter that the bootstrap CI on
$A$ summarises only implicitly; a cell with a large
$\sigma_{\log_{10} A}$ has a wide bootstrap CI on $A$ for the same
reason. Second, the fixed-effect slope $\hat{k}_{\mathrm{LME}}$ with
its $1.96\,\mathrm{SE}$ Wald interval is the parametric analogue of
the main-table $\hat{k}_{\mathrm{OLS}}$ with its bootstrap percentile
CI; substantive agreement between the two estimators is the
specific cross-check this appendix asks.
Table~\ref{tab:mixed-effects} reports the two side by side for
every populated cell. A reader who finds the LME slope outside the
bootstrap CI on any cell should treat the main estimator as
authoritative --- the percentile CI bounds the satellite-level
resampling distribution directly, while the LME fixed-effect SE
assumes a normal random-intercept distribution that thin or
unbalanced cells violate.

\begin{table}[ht]
  \centering
  \caption{Per-cell comparison of the main-text per-pair OLS slope
  $\hat{k}_{\mathrm{OLS}}$ (with $95\%$ satellite-level bootstrap
  percentile CI, identical procedure to Table~\ref{tab:powerlaw})
  against the linear mixed-effects fixed-effect slope
  $\hat{k}_{\mathrm{LME}}$ (with $1.96\,\mathrm{SE}$ Wald CI). The
  random-intercept SD $\hat\sigma_{\log_{10} A}$ is the LME estimate
  of the per-satellite scatter in $\log_{10} A$ that the bootstrap
  on $A$ absorbs implicitly. One row per
  (altitude shell $\times$ pooled generation $\times$ propagator)
  cell that reaches the convergence guardrails
  ($\ge 30$ pairs and $\ge 5$ satellites in the cell).}
  \label{tab:mixed-effects}
  \resizebox{\textwidth}{!}{\input{tables/tab_mixed_effects.tex}}
\end{table}

\section*{Data and code availability}
\label{sec:availability}

All code is available at
\url{https://github.com/astro-tools/paper-tle-divergence-atlas}
under the MIT license. The full sweep output bundle is archived on
Zenodo at \url{https://doi.org/10.5281/zenodo.20277028} (concept DOI,
resolving to the latest version); the bundle contains the
$24{,}641$-pair NRLMSISE-00 main sweep, the $C_{D}\cdot A \times
\{0.8,\,1.2\}$ and $\{50,\,200\}$\,m maneuver-threshold
sensitivity-subset outputs, the truth-floor diagnostic frame,
the mixed-effects supplement, the GMAT mission script, the
resumable sweep manifest, and the EGM2008 installer. Starlink TLE
data is publicly available from the
\href{https://www.space-track.org}{Space-Track \texttt{gp\_history}
endpoint}; daily solar-activity indices are from
\href{https://celestrak.org/SpaceData/}{CelesTrak's space-weather file}.

\bibliographystyle{plainnat}
\bibliography{bib}

\end{document}

%% file: tables/tab_powerlaw.tex
\begin{tabular}{llrrrrrrrr}
\toprule
 & & \multicolumn{4}{c}{SGP4} & \multicolumn{4}{c}{high-fid} \\
\cmidrule(lr){3-6} \cmidrule(lr){7-10}
shell & generation & $A$ [95\% CI] & $k$ [95\% CI] & $R^{2}$ & $p_{k=1}/p_{k=2}$ & $A$ [95\% CI] & $k$ [95\% CI] & $R^{2}$ & $p_{k=1}/p_{k=2}$ \\
\midrule
540 km & v1.x & 0.19 [0.17, 0.2] & 0.84 [0.82, 0.87] & 0.40 & $<\!10^{-3}$ / $<\!10^{-3}$ & \textbf{0.13 [0.12, 0.14]} & 1.28 [1.26, 1.30] & 0.68 & $<\!10^{-3}$ / $<\!10^{-3}$ \\
550 km & v1.x & 0.23 [0.19, 0.28] & 0.75 [0.69, 0.83] & 0.33 & $<\!10^{-3}$ / $<\!10^{-3}$ & \textbf{0.23 [0.19, 0.27]} & 0.99 [0.95, 1.04] & 0.55 & 0.760 / $<\!10^{-3}$ \\
550 km & v2-mini & \textbf{0.049 [0.045, 0.053]} & 1.44 [1.42, 1.46] & 0.77 & $<\!10^{-3}$ / $<\!10^{-3}$ & 0.15 [0.13, 0.18] & 1.15 [1.10, 1.19] & 0.67 & $<\!10^{-3}$ / $<\!10^{-3}$ \\
560 km & v1.x & \textbf{0.15 [0.13, 0.17]} & 0.86 [0.82, 0.89] & 0.41 & $<\!10^{-3}$ / $<\!10^{-3}$ & 0.15 [0.13, 0.17] & 1.12 [1.08, 1.15] & 0.58 & $<\!10^{-3}$ / $<\!10^{-3}$ \\
560 km & v2-mini & \textbf{0.073 [0.066, 0.083]} & 1.32 [1.29, 1.35] & 0.73 & $<\!10^{-3}$ / $<\!10^{-3}$ & 0.18 [0.16, 0.2] & 1.15 [1.11, 1.18] & 0.66 & $<\!10^{-3}$ / $<\!10^{-3}$ \\
\bottomrule
\end{tabular}

%% file: tables/tab_propagator_wins.tex
\begin{tabular}{llrrr}
\toprule
Shell & $\Delta t$ & $n_{\mathrm{pairs}}$ & \multicolumn{1}{c}{hi-fid wins, 3D L$_{2}$ [95\% CI]} & \multicolumn{1}{c}{hi-fid wins, along-track [95\% CI]} \\
(km) & & & & \\
\midrule
540 & 6h & 1,957 & 34.6\% [32.3, 36.8] & 35.6\% [33.2, 37.8] \\
540 & 1d & 2,854 & 17.6\% [16.2, 19.1] & 17.5\% [16.1, 19.1] \\
540 & 3d & 1,748 & 6.5\% [5.4, 8.0] & 6.5\% [5.4, 8.0] \\
540 & 7d & 920 & 5.4\% [3.9, 7.0] & 5.4\% [3.9, 7.0] \\
\midrule
550 & 6h & 2,031 & 39.8\% [37.0, 42.6] & 42.8\% [39.9, 45.8] \\
550 & 1d & 3,351 & 30.3\% [28.8, 31.9] & 30.4\% [28.9, 31.9] \\
550 & 3d & 2,587 & 45.1\% [42.0, 47.9] & 45.1\% [42.0, 47.9] \\
550 & 7d & 1,562 & 60.2\% [55.9, 64.4] & 60.2\% [55.9, 64.4] \\
\midrule
560 & 6h & 1,433 & 23.9\% [21.3, 26.5] & 26.8\% [24.1, 29.4] \\
560 & 1d & 2,725 & 26.4\% [24.5, 28.4] & 26.3\% [24.4, 28.4] \\
560 & 3d & 2,124 & 27.0\% [24.3, 30.1] & 26.9\% [24.2, 30.1] \\
560 & 7d & 1,349 & 26.7\% [22.1, 31.6] & 26.6\% [21.9, 31.6] \\
\midrule
\textit{pooled} & 6h & 5,421 & 33.7\% [32.2, 35.3] & 36.0\% [34.3, 37.6] \\
\textit{pooled} & 1d & 8,930 & 25.1\% [23.9, 26.1] & 25.0\% [23.9, 26.1] \\
\textit{pooled} & 3d & 6,459 & 28.7\% [26.6, 30.9] & 28.7\% [26.6, 30.9] \\
\textit{pooled} & 7d & 3,831 & 35.3\% [31.9, 38.8] & 35.2\% [31.9, 38.8] \\
\bottomrule
\end{tabular}

%% file: tables/tab_propagator_wins_by_gen.tex
\begin{tabular}{lllrrr}
\toprule
Shell & $\Delta t$ & Cohort & $n_{\mathrm{pairs}}$ & \multicolumn{1}{c}{hi-fid wins, 3D L$_{2}$ [95\% CI]} & \multicolumn{1}{c}{hi-fid wins, along-track [95\% CI]} \\
(km) & & & & & \\
\midrule
540 & 6h & v1.x & 1,957 & 34.6\% [32.3, 36.8] & 35.6\% [33.2, 37.8] \\
 & 1d & v1.x & 2,854 & 17.6\% [16.2, 19.1] & 17.5\% [16.1, 19.1] \\
 & 3d & v1.x & 1,748 & 6.5\% [5.4, 8.0] & 6.5\% [5.4, 8.0] \\
 & 7d & v1.x & 920 & 5.4\% [3.9, 7.0] & 5.4\% [3.9, 7.0] \\
\midrule
550 & 6h & v1.x & 732 & 31.0\% [27.7, 34.6] & 34.8\% [31.1, 39.0] \\
 &  & v2-mini & 1,299 & 44.7\% [41.6, 47.6] & 47.3\% [43.8, 50.7] \\
 & 1d & v1.x & 916 & 26.6\% [24.5, 28.8] & 26.7\% [24.6, 28.8] \\
 &  & v2-mini & 2,435 & 31.7\% [29.9, 34.0] & 31.7\% [29.9, 33.9] \\
 & 3d & v1.x & 634 & 16.6\% [13.4, 20.0] & 16.6\% [13.4, 20.0] \\
 &  & v2-mini & 1,953 & 54.3\% [52.4, 56.1] & 54.3\% [52.4, 56.1] \\
 & 7d & v1.x & 341 & 16.1\% [11.0, 22.2] & 16.1\% [11.0, 22.2] \\
 &  & v2-mini & 1,221 & 72.6\% [69.9, 75.4] & 72.6\% [69.9, 75.4] \\
\midrule
560 & 6h & v1.x & 814 & 29.0\% [25.9, 32.0] & 31.7\% [28.2, 34.9] \\
 &  & v2-mini & 619 & 17.1\% [14.0, 20.6] & 20.4\% [17.1, 23.8] \\
 & 1d & v1.x & 1,659 & 23.6\% [21.6, 25.5] & 23.5\% [21.5, 25.5] \\
 &  & v2-mini & 1,066 & 30.8\% [27.0, 34.8] & 30.8\% [27.0, 34.9] \\
 & 3d & v1.x & 1,333 & 17.2\% [14.9, 19.4] & 17.1\% [14.8, 19.4] \\
 &  & v2-mini & 791 & 43.5\% [39.1, 47.6] & 43.5\% [39.1, 47.6] \\
 & 7d & v1.x & 888 & 11.9\% [8.8, 15.5] & 11.8\% [8.7, 15.4] \\
 &  & v2-mini & 461 & 55.1\% [49.9, 60.4] & 55.1\% [49.9, 60.4] \\
\bottomrule
\end{tabular}

%% file: tables/tab_maneuver_rejections.tex
\begin{tabular}{llrrrr}
\toprule
Shell & $\Delta t$ & Candidates & Survivors & Rejected & Rejected (\%) \\
(km) & & & & & \\
\midrule
540 & 6 h & 2{,}019 & 1{,}957 & 62 & 3.1\% \\
540 & 1 d & 3{,}030 & 2{,}854 & 176 & 5.8\% \\
540 & 3 d & 2{,}003 & 1{,}748 & 255 & 12.7\% \\
540 & 7 d & 1{,}188 & 920 & 268 & 22.6\% \\
\midrule
550 & 6 h & 2{,}239 & 2{,}031 & 208 & 9.3\% \\
550 & 1 d & 3{,}800 & 3{,}351 & 449 & 11.8\% \\
550 & 3 d & 3{,}323 & 2{,}587 & 736 & 22.1\% \\
550 & 7 d & 2{,}384 & 1{,}562 & 822 & 34.5\% \\
\midrule
560 & 6 h & 1{,}615 & 1{,}433 & 182 & 11.3\% \\
560 & 1 d & 3{,}301 & 2{,}725 & 576 & 17.4\% \\
560 & 3 d & 3{,}043 & 2{,}124 & 919 & 30.2\% \\
560 & 7 d & 2{,}506 & 1{,}349 & 1{,}157 & 46.2\% \\
\midrule
Total & & 30{,}451 & 24{,}641 & 5{,}810 & 19.1\% \\
\bottomrule
\end{tabular}

%% file: tables/tab_cda_sensitivity.tex
\begin{tabular}{llrrr}
\toprule
Shell & $\Delta t$ & $\mathrm{med}\,|\Delta\mathbf{r}|_{\mathrm{hifi}}$ (1.0$\times$) & 0.8$\times$ shift & 1.2$\times$ shift \\
(km) & & (km) & (vs.\ baseline) & (vs.\ baseline) \\
\midrule
550 & 6 h & 1.29 & -1.4\% & +1.4\% \\
550 & 1 d & 6.60 & -0.4\% & -2.4\% \\
550 & 3 d & 23.35 & -9.1\% & +2.1\% \\
550 & 7 d & 77.20 & -9.4\% & +5.8\% \\
\midrule
560 & 6 h & 1.76 & +0.2\% & +0.7\% \\
560 & 1 d & 8.12 & -2.4\% & +2.3\% \\
560 & 3 d & 31.47 & -4.8\% & +4.8\% \\
560 & 7 d & 72.38 & -12.4\% & +12.4\% \\
\bottomrule
\end{tabular}

%% file: tables/tab_maneuver_threshold.tex
\begin{tabular}{llrrr}
\toprule
Shell & $\Delta t$ & $\mathrm{med}\,|\Delta\mathbf{r}|_{\mathrm{hifi}}$ (100 m, 95\% CI) & 50 m shift & 200 m shift \\
(km) & & (km) & (vs.\ baseline) & (vs.\ baseline) \\
\midrule
540 & 6 h & 1.57 [1.43, 1.69] & -5.5\% & +2.2\% \\
540 & 1 d & 8.90 [8.34, 9.46] & -4.8\% & +1.3\% \\
540 & 3 d & 36.40 [33.73, 38.14] & -5.7\% & +1.2\% \\
540 & 7 d & 110.80 [103.54, 115.42] & -6.9\%\textsuperscript{$\dagger$} & +1.3\% \\
\midrule
550 & 6 h & 1.50 [1.37, 1.60] & -7.1\% & +3.9\% \\
550 & 1 d & 6.40 [6.06, 6.74] & -5.1\% & +3.6\% \\
550 & 3 d & 22.00 [21.36, 22.82] & -4.1\%\textsuperscript{$\dagger$} & +1.9\% \\
550 & 7 d & 68.55 [65.05, 72.44] & -3.0\% & +0.6\% \\
\midrule
560 & 6 h & 1.43 [1.34, 1.52] & -4.7\% & +3.5\% \\
560 & 1 d & 6.14 [5.58, 6.75] & -5.7\% & +4.4\% \\
560 & 3 d & 21.86 [20.12, 23.89] & -3.3\% & +0.6\% \\
560 & 7 d & 69.87 [65.53, 75.53] & -0.1\% & +0.6\% \\
\bottomrule
\end{tabular}

%% file: tables/tab_mixed_effects.tex
\begin{tabular}{llcrrrr}
\toprule
shell & generation & propagator & $\hat k_{\mathrm{OLS}}$ [95\% CI] & $\hat k_{\mathrm{LME}}$ [95\% CI] & $\hat\sigma_{\log_{10}A}$ & $n_{\mathrm{pairs}}$ \\
\midrule
540 km & v1.x & SGP4 & 0.842 [0.818, 0.868] & 0.842 [0.819, 0.865] & 0.10 & 7479 \\
540 km & v1.x & high-fid & 1.28 [1.26, 1.3] & 1.26 [1.24, 1.28] & 0.09 & 7479 \\
550 km & v1.x & SGP4 & 0.752 [0.691, 0.825] & 0.757 [0.717, 0.798] & 0.11 & 2623 \\
550 km & v1.x & high-fid & 0.995 [0.949, 1.04] & 1 [0.967, 1.04] & 0.08 & 2623 \\
550 km & v2-mini & SGP4 & 1.44 [1.42, 1.46] & 1.44 [1.42, 1.46] & 0.04 & 6908 \\
550 km & v2-mini & high-fid & 1.15 [1.1, 1.19] & 1.15 [1.13, 1.17] & 0.04 & 6908 \\
560 km & v1.x & SGP4 & 0.857 [0.823, 0.891] & 0.856 [0.827, 0.885] & 0.13 & 4694 \\
560 km & v1.x & high-fid & 1.12 [1.08, 1.15] & 1.1 [1.07, 1.13] & 0.12 & 4694 \\
560 km & v2-mini & SGP4 & 1.32 [1.29, 1.35] & 1.32 [1.29, 1.35] & 0.04 & 2937 \\
560 km & v2-mini & high-fid & 1.15 [1.11, 1.18] & 1.13 [1.1, 1.16] & 0.10 & 2937 \\
\bottomrule
\end{tabular}

%% file: ms.bbl
\begin{thebibliography}{22}
\providecommand{\natexlab}[1]{#1}
\providecommand{\url}[1]{\texttt{#1}}
\expandafter\ifx\csname urlstyle\endcsname\relax
  \providecommand{\doi}[1]{doi: #1}\else
  \providecommand{\doi}{doi: \begingroup \urlstyle{rm}\Url}\fi

\bibitem[Acciarini et~al.(2025)Acciarini, Baydin, and
  Izzo]{acciarini_dsgp4_2025}
Giacomo Acciarini, At\i{}l\i{}m~G\"{u}ne\c{s} Baydin, and Dario Izzo.
\newblock Closing the gap between {SGP4} and high-precision propagation via
  differentiable programming.
\newblock \emph{Acta Astronautica}, 226:\penalty0 577--585, 2025.
\newblock \doi{10.1016/j.actaastro.2024.10.063}.
\newblock arXiv:2402.04830.

\bibitem[Baruah et~al.(2024)Baruah, Roy, Sinha, Palmerio, Pal, Oliveira, and
  Nandy]{baruah2024}
Yoshita Baruah, Souvik Roy, Suvadip Sinha, Erika Palmerio, Sanchita Pal,
  Denny~M. Oliveira, and Dibyendu Nandy.
\newblock The {L}oss of {S}tarlink {S}atellites in {F}ebruary 2022: {H}ow
  {M}oderate {G}eomagnetic {S}torms {C}an {A}dversely {A}ffect {A}ssets in
  {L}ow-{E}arth {O}rbit.
\newblock \emph{Space Weather}, 22\penalty0 (4):\penalty0 e2023SW003716, 2024.
\newblock \doi{10.1029/2023SW003716}.

\bibitem[Bowman et~al.(2008)Bowman, Tobiska, Marcos, Huang, Lin, and
  Burke]{bowman2008}
Bruce~R. Bowman, W.~Kent Tobiska, Frank~A. Marcos, Cheryl~Y. Huang, Chin~S.
  Lin, and William~J. Burke.
\newblock A new empirical thermospheric density model {JB2008} using new solar
  and geomagnetic indices.
\newblock In \emph{AIAA/AAS Astrodynamics Specialist Conference}, August 2008.
\newblock \doi{10.2514/6.2008-6438}.
\newblock AIAA 2008-6438.

\bibitem[Emmert et~al.(2021)Emmert, Drob, Picone, Siskind, Jones, Mlynczak,
  Bernath, Chu, Doornbos, Funke, Goncharenko, Hervig, Schwartz, Sheese, Vargas,
  Williams, and Yuan]{emmert2021}
J.~T. Emmert, D.~P. Drob, J.~M. Picone, D.~E. Siskind, Jr. Jones, M., M.~G.
  Mlynczak, P.~F. Bernath, X.~Chu, E.~Doornbos, B.~Funke, L.~P. Goncharenko,
  M.~E. Hervig, M.~J. Schwartz, P.~E. Sheese, F.~Vargas, B.~P. Williams, and
  T.~Yuan.
\newblock {NRLMSIS} 2.0: {A} whole-atmosphere empirical model of temperature
  and neutral species densities.
\newblock \emph{Earth and Space Science}, 8\penalty0 (3):\penalty0
  e2020EA001321, 2021.
\newblock \doi{10.1029/2020EA001321}.

\bibitem[{eoPortal}(2022)]{leolabs_eoportal}
{eoPortal}.
\newblock {LeoLabs} commercial ground-based tracking service for {LEO} resident
  space objects.
\newblock ESA eoPortal Directory;
  \url{https://www.eoportal.org/ftp/satellite-missions/l/LeoLabs_070122/LeoLabs.html},
  2022.
\newblock Accessed 2026-05-14.

\bibitem[Hoots et~al.(2004)Hoots, Schumacher, and Glover]{hoots2004}
Felix~R. Hoots, Paul~W. Schumacher, and Robert~A. Glover.
\newblock History of analytical orbit modeling in the {U}.\,{S}. {S}pace
  {S}urveillance {S}ystem.
\newblock \emph{Journal of Guidance, Control, and Dynamics}, 27\penalty0
  (2):\penalty0 174--185, 2004.
\newblock \doi{10.2514/1.9161}.

\bibitem[Krebs(2024)]{krebs_starlink_v2mini}
Gunter~D. Krebs.
\newblock {S}tarlink {B}lock v2-{M}ini.
\newblock Gunter's Space Page;
  \url{https://space.skyrocket.de/doc_sdat/starlink-v2-mini.htm}, 2024.
\newblock Accessed 2026-05-11.

\bibitem[Lang and Jiang(2025)]{lang2025}
Anqi Lang and Yu~Jiang.
\newblock Orbit {D}etermination for {C}ontinuously {M}aneuvering {S}tarlink
  {S}atellites {B}ased on an {U}nscented {B}atch {F}iltering {M}ethod.
\newblock \emph{Sensors}, 25\penalty0 (13):\penalty0 4079, 2025.
\newblock \doi{10.3390/s25134079}.

\bibitem[Lemmens and Krag(2014)]{lemmens2014}
Stijn Lemmens and Holger Krag.
\newblock Two-{L}ine-{E}lements-{B}ased {M}aneuver {D}etection {M}ethods for
  {S}atellites in {L}ow {E}arth {O}rbit.
\newblock \emph{Journal of Guidance, Control, and Dynamics}, 37\penalty0
  (3):\penalty0 860--868, 2014.
\newblock \doi{10.2514/1.61300}.

\bibitem[McDowell(2020)]{mcdowell2020}
Jonathan~C. McDowell.
\newblock The {L}ow {E}arth {O}rbit {S}atellite {P}opulation and {I}mpacts of
  the {S}pace{X} {S}tarlink {C}onstellation.
\newblock \emph{The Astrophysical Journal Letters}, 892\penalty0 (2):\penalty0
  L36, 2020.
\newblock \doi{10.3847/2041-8213/ab8016}.
\newblock GCAT data product at \url{https://planet4589.org/space/gcat/}.

\bibitem[Pavlis et~al.(2012)Pavlis, Holmes, Kenyon, and Factor]{pavlis2012}
Nikolaos~K. Pavlis, Simon~A. Holmes, Steve~C. Kenyon, and John~K. Factor.
\newblock The development and evaluation of the {E}arth {G}ravitational {M}odel
  2008 ({EGM2008}).
\newblock \emph{Journal of Geophysical Research: Solid Earth}, 117\penalty0
  (B4):\penalty0 B04406, 2012.
\newblock \doi{10.1029/2011JB008916}.

\bibitem[Payne et~al.(2022)Payne, Hoots, Butkus, Slatton, and
  Nguyen]{payne_sgp4xp_2022}
Timothy Payne, Felix Hoots, Albert Butkus, Zachary Slatton, and Dinh Nguyen.
\newblock Improvements to the {SGP4} propagator ({SGP4-XP}).
\newblock In \emph{Proceedings of the Advanced {M}aui {O}ptical and {S}pace
  {S}urveillance {T}echnologies ({AMOS}) Conference}, 2022.
\newblock URL
  \url{https://amostech.com/TechnicalPapers/2022/Astrodynamics/Payne_2.pdf}.

\bibitem[Petit and Luzum(2010)]{iers2010}
G{\'e}rard Petit and Brian Luzum.
\newblock {IERS} {C}onventions (2010).
\newblock IERS Technical Note~36, International Earth Rotation and Reference
  Systems Service, Frankfurt am Main, 2010.
\newblock Verlag des Bundesamts f{\"u}r Kartographie und Geod{\"a}sie, ISBN
  3-89888-989-6.

\bibitem[Picone et~al.(2002)Picone, Hedin, Drob, and Aikin]{picone2002}
J.~M. Picone, A.~E. Hedin, D.~P. Drob, and A.~C. Aikin.
\newblock {NRLMSISE-00} empirical model of the atmosphere: {S}tatistical
  comparisons and scientific issues.
\newblock \emph{Journal of Geophysical Research: Space Physics}, 107\penalty0
  (A12):\penalty0 1468, 2002.
\newblock \doi{10.1029/2002JA009430}.

\bibitem[Rhodes(2024)]{rhodes_sgp4}
Brandon Rhodes.
\newblock {sgp4}: {P}ython implementation of the {SGP4}/{SDP4}
  satellite-tracking algorithm.
\newblock Python package; \url{https://github.com/brandon-rhodes/python-sgp4},
  2024.
\newblock Implementation of the reference {SGP4} code from \citet{vallado2006};
  accessed 2026-05-13.

\bibitem[{Space Exploration Holdings, LLC}(2024)]{spacex_fcc_semiannual_2024}
{Space Exploration Holdings, LLC}.
\newblock {S}emi-{A}nnual {C}onstellation {S}tatus {R}eport, 1~{D}ecember 2023
  to 31~{M}ay 2024.
\newblock Federal Communications Commission filing, IBFS File
  No.~SATMOD2020041700037, 2024.
\newblock Filed 1~July 2024. Reports an average of 14 thruster firings per
  satellite over the six-month window for autonomous collision-avoidance
  maneuvers.

\bibitem[{Space Exploration Holdings, LLC}(2025)]{spacex_fcc_pnt}
{Space Exploration Holdings, LLC}.
\newblock Reply comments on the {N}otice of {I}nquiry, {WT} {D}ocket
  {N}o.~25-110: {P}romoting the {D}evelopment of {P}ositioning, {N}avigation,
  and {T}iming {T}echnologies and {S}olutions.
\newblock Federal Communications Commission filing, 2025.

\bibitem[Vallado(2013)]{vallado2013}
David~A. Vallado.
\newblock \emph{Fundamentals of Astrodynamics and Applications}.
\newblock Space Technology Library. Microcosm Press, Hawthorne, CA, 4th
  edition, 2013.
\newblock ISBN 978-1881883180.

\bibitem[Vallado and Cefola(2012)]{vallado2012}
David~A. Vallado and Paul~J. Cefola.
\newblock Two-line {E}lement {S}ets --- {P}ractice and {U}se.
\newblock In \emph{63rd International Astronautical Congress}, 2012.
\newblock IAC-12-A6.6.11.

\bibitem[Vallado and Crawford(2008)]{vallado_crawford2008}
David~A. Vallado and Paul Crawford.
\newblock {SGP4} {O}rbit {D}etermination.
\newblock In \emph{AIAA/AAS Astrodynamics Specialist Conference and Exhibit},
  August 2008.
\newblock \doi{10.2514/6.2008-6770}.
\newblock AIAA 2008-6770.

\bibitem[Vallado et~al.(2006)Vallado, Crawford, Hujsak, and Kelso]{vallado2006}
David~A. Vallado, Paul Crawford, Richard Hujsak, and T.~S. Kelso.
\newblock Revisiting {S}pacetrack {R}eport \#3.
\newblock In \emph{AIAA/AAS Astrodynamics Specialist Conference and Exhibit},
  August 2006.
\newblock \doi{10.2514/6.2006-6753}.
\newblock AIAA 2006-6753.

\bibitem[Verner(1978)]{verner1978}
J.~H. Verner.
\newblock Explicit {R}unge--{K}utta methods with estimates of the local
  truncation error.
\newblock \emph{SIAM Journal on Numerical Analysis}, 15\penalty0 (4):\penalty0
  772--790, 1978.
\newblock \doi{10.1137/0715051}.

\end{thebibliography}
